\documentclass[twocolumn,amssymb, nobibnotes, aps, prd,amsmath,mathtools, nofootinbib,pgfmath,pgffor,superscriptaddress]{revtex4-1}
\usepackage{graphicx}
\usepackage[normalem]{ulem}
\usepackage{amsthm}

\newcommand{\be}{\begin{equation}}
\newcommand{\ee}{\end{equation}}
\usepackage{caption,color}
\usepackage{subcaption}
\usepackage{multirow}

\usepackage{hyperref}
\hypersetup{colorlinks,citecolor=blue}

\usepackage[dvipsnames]{xcolor}
\usepackage{color, colortbl}
\usepackage{bigints}

\begin{document}

\title[Which black hole modes ring freely? Horizon regularity, the forced ringdown transient, and GW250114]{Which black hole modes ring freely? Horizon regularity, the forced ringdown transient, and GW250114\\[0.8ex] {\normalsize\normalfont\itshape Dedicated to the memory of Rafael Sorkin who inspired this work}}

\author{Jahed Abedi}
\email{jahed.abedi@bimsa.cn}
\affiliation{Beijing Institute of Mathematical Sciences and Applications (BIMSA), Huairou District, Beijing 101408, P. R. China}

\author{Niayesh Afshordi}
\email[]{nafshordi@pitp.ca }
\affiliation{Department of Physics and Astronomy, University of Waterloo,
200 University Ave W, N2L 3G1, Waterloo, Canada
}
\affiliation{Waterloo Centre for Astrophysics, University of Waterloo, Waterloo, ON, N2L 3G1, Canada}
\affiliation{Perimeter Institute For Theoretical Physics, 31 Caroline St N, Waterloo, Canada}

\begin{abstract}
The quasinormal modes (QNMs) of a black hole are defined by keeping only ingoing waves at the horizon. We ask what regularity at the horizon requires of the outgoing waves that this condition discards. For a scalar-like perturbation, bounded first derivatives in a frame regular at the future horizon rule out outgoing waves that decay more slowly than the surface gravity $\kappa$, and smoothness singles out a ladder of frequencies, $\omega=m\Omega_{\rm H}-ij\kappa$ with $j=1,2,\dots$, where $\Omega_{\rm H}$ is the angular velocity of the horizon. We call these the horizon modes. For Schwarzschild black holes and for the dominant $\ell=m=2$ modes of Kerr black holes, only the fundamental QNM decays more slowly than $\kappa$. These local statements leave the QNM spectrum unchanged, but they matter for the ringdown of a merger, because a source falling through the horizon drives the exterior at the horizon frequencies. Each QNM then has a driven part and a free part, and the free part takes over only once the driving decays faster than the mode. The fundamental mode separates from the driving early. The overtones stay driven until the source is hidden behind the potential barrier, so overtone amplitudes fitted near the peak of the signal are in general not amplitudes of free oscillations. In GW250114, a ringdown model made of the fundamental mode and the first horizon mode, $\omega=2\Omega_{\rm H}-i\kappa$, is preferred over the fundamental mode plus its first overtone, with Bayes factors between 2 and 30, for start times from $2M_f$ before to $8M_f$ after coalescence. The horizon-mode fit returns a remnant that is heavier and spins faster than the inspiral predicts. Such a shift is expected from nonlinear backreaction, because the horizon that emits the mode has not yet settled to its final mass and spin. Zero-noise injections of waveforms calibrated to numerical relativity reproduce both the preference and the shift, which are therefore features of general relativity.
\end{abstract}

\maketitle

\section{\label{Introduction}Introduction}

In general relativity the event horizon of a black hole is the boundary of the region from which signals can reach distant observers. It is a global property of the space-time, and from the point of view of an asymptotic observer an infalling wavepacket takes an infinite time to cross it. Linear scalar, vector, or tensor perturbations around a stationary black hole satisfy a generalized wave equation. They become important in the final stage of a binary black hole merger, when the remnant ``rings down'' to a stationary solution. The backreaction of the perturbations enters at second order, through a stress (or energy-momentum) tensor that is quadratic in first derivatives of the linear solution \cite{PhysRev.166.1272}. Late in the ringdown the signal is usually described as a superposition of quasinormal modes (QNMs): damped oscillations with complex frequencies $\omega=\omega_{R}-i\omega_{I}$, found by solving the linear wave equation with an outgoing boundary condition at infinity and an ingoing one at the horizon \cite{Kokkotas:1999bd} (see also Fig.~5 of \cite{TheLIGOScientific:2016src}).

In this paper we ask which regularity condition a perturbation must satisfy at the horizon, and what that condition selects. Our starting point is that the backreaction of the perturbation has to stay finite as the horizon is approached, since otherwise linear perturbation theory is not consistent there. Finiteness conditions have been used before to give a geometric formulation of the QNM boundary conditions \cite{Zenginoglu:2011jz,PanossoMacedo:2018hab}. Here we apply the idea to the stress tensor of the gravitational waves themselves, i.e., to the second-order backreaction of the perturbations on the metric. QNMs can also be defined as poles of a meromorphic extension of the Green's function of the wave equation (e.g., \cite{Gajic:2019qdd,Gajic:2019oem}), and which of them appear as isolated eigenvalues, and not as part of a continuum, is known to depend on the regularity of the function space in which the problem is posed~\cite{Warnick:2013hr,Gajic:2019oem}.

As we shall see, generic outgoing solutions of the perturbation equations are not differentiable at the future horizon. We have to check whether this is a physical problem or a coordinate artifact. Finite backreaction translates into finite first derivatives in a locally inertial frame near the horizon, which is a weaker requirement than the standard QNM boundary condition. We shall see that the backreaction of an outgoing wave diverges at the horizon only if the wave is damped slowly, with $\omega_I<\kappa$, where $\kappa$ is the surface gravity.

For Schwarzschild black holes, and for the Kerr modes examined in Sec.~\ref{sec:forced_kerr}, the only QNM in this range is the fundamental one. Regularity thus enforces the ingoing boundary condition for the fundamental mode, but it does not distinguish the overtones from other outgoing solutions whose derivatives are bounded at the horizon. This is a statement about what is locally admissible. It does not remove the overtones from the QNM spectrum of the retarded problem, which is fixed by causal boundary conditions. 
If we also require the solution to be smooth ($\mathcal{C}^{\infty}$) in the local inertial frame, the outgoing waves are confined to a ``quantized'' set, $\omega_{I} = j\kappa$ and $\omega_{R}=0$ with $j \in \mathbb{N}$ (for a Kerr black hole, $\omega_R=m\Omega_{\rm H}$, with $\Omega_{\rm H}$ the angular velocity of the horizon). We call the members of this set horizon modes. They can emerge from the backreaction of ingoing gravitational waves, or from infalling matter, during the merger: we show that they are the frequencies at which a smooth source falling through the horizon acts on the exterior. They do not reach infinity as stationary spectral lines, because their residues cancel in the waveform~\cite{Zimmerman2011HorizonMode,Kuntz2026RedshiftModes,Kubota2026HorizonCancel}, but they organize the forcing of the ringdown while the source approaches the horizon.

Which of the damped sinusoids in a ringdown signal are free oscillations of the remnant has become a pressing question. Fits of numerical-relativity (NR) waveforms with many overtones, starting at the peak of the signal, recover the remnant accurately and give stable overtone amplitudes~\cite{Giesler:2019uxc,Giesler2025OvertonesNonlin}, and a Bayesian analysis of high-accuracy waveforms finds overtones, retrograde modes and nonlinear modes~\cite{Dyer2025ModeContent}. Other studies find that high overtones can absorb initial-data effects, nonlinearities and tails~\cite{Baibhav2023Agnostic,Nee2023OvertoneRole,Cheung2023Nonlinear,Mitman2023Nonlinear}, that in wave-packet scattering the branch-cut contribution can reach about ten per cent of the QNM signal at the onset of the ringdown~\cite{GuptaOshita2026EarlyTails}, and that the overtones are far more sensitive than the fundamental mode to small perturbations of the wave operator~\cite{Jaramillo2021Pseudospectrum}. In parallel, filtering the dominant QNMs out of merger waveforms leaves a residual, called the ``direct wave'', that has been attributed to the plunging source near the horizon~\cite{Oshita2025DirectWaves,MaWang2026FoundationsDW,Sun2026DWComplexFreq} and reported in GW250114~\cite{Lu2026GW250114Horizon,Chung2026DWAreaLaw}. How much the direct wave says about the horizon is debated. The horizon-frequency poles of a plunging source cancel in the signal at infinity, in Schwarzschild~\cite{Kuntz2026RedshiftModes} and in Kerr~\cite{Kubota2026HorizonCancel}. The fitted frequency and damping of the residual do not follow the horizon values in general~\cite{Kankani2026NotMeaningful,Dyer2026ModelingDW,Han2026RedshiftTransfer}. And the most recent analyses describe the direct wave as a source-driven signal whose frequency sweeps through the plunge and is screened by the potential barrier, not as a fixed horizon line~\cite{Nakamoto2026PoleSplitting,Laeuger2026SourceImprints}. Meanwhile, for a particle plunging into a Kerr black hole, each QNM contribution has been shown to stay driven by the source until the source decays faster than the mode~\cite{Weller2026DrivenToFree}, and the time-dependent excitation of QNMs has been formulated in general~\cite{DeAmicis2026DynQNM,DeAmicis2026DynQNMII}.

We argue that these results fit together once the ringdown is treated as the forced response to a source that crosses the horizon. The regularity threshold $\omega_I=\kappa$ of Sec.~\ref{sec:regularity} is the rate at which such a source settles onto the horizon, and the horizon modes are the frequencies at which it acts. The fundamental mode decays more slowly than this forcing and separates from it early. An overtone decays faster than the forcing until the source is screened by the potential barrier, so it is driven during the first few $M$ of the ringdown. A damped sinusoid at the first horizon frequency is then a natural template for the driven part of the signal, and we test it on waveform models calibrated to numerical relativity simulations of black hole mergers, as well as on GW250114, the merger with the highest signal-to-noise ratio reported by the LIGO--Virgo--KAGRA Collaboration~\cite{LVK2025GW250114}. Throughout, we keep three kinds of statement apart: local regularity, the analytic structure of the retarded response, and what can be identified in finite, noisy data.

The paper is organized as follows. Section~\ref{sec:regularity} derives the regularity threshold and the ladder of horizon modes, for Schwarzschild and Kerr black holes. Section~\ref{sec:forced} develops the forced-ringdown picture, states which modes ring freely and when, and lists its testable predictions. Section~\ref{sec:horizon} confronts it with GW250114, and Sec.~\ref{Conclusion} concludes. We use units $G=c=1$. Time dependence is $e^{-i\omega t}$ with $\omega=\omega_R-i\omega_I$, so that $\omega_I>0$ is a decay rate; for the $n$-th QNM we also write $\omega_n=\Omega_n-i\gamma_n$.

\section{\label{sec:regularity}Horizon regularity and the spectrum of outgoing modes}
\subsection{\label{Backreaction limit on quasinormal modes} Backreaction limit on quasinormal modes}
A realistic black hole is perturbed from outside, by an object or a wave that falls through the future horizon. To follow the infall we need coordinates that are regular at the future horizon. Near the horizon they are related to the static coordinates in the way Minkowski coordinates are related to those of a Rindler observer, and nothing in the solution diverges in them as $t\to\infty$.

Near the horizon we therefore use ingoing and outgoing plane waves in Rindler coordinates for a scalar field, which we take as a model for a metric perturbation around Minkowski space in the TT gauge (written in Rindler coordinates). In this model, we assume the perturbation is a scalar-like quantity whose first derivatives in a regular frame enter the backreaction, as in Isaacson's effective stress tensor~\cite{PhysRev.166.1272}.  Isaacson's tensor relies on a short-wavelength average, and for a tensor perturbation the second-order Einstein tensor contains terms of the forms $\nabla h\nabla h$ and $h\nabla\nabla h$, which do not reduce to a unique local stress built from first derivatives alone. The threshold we derive is thus a scalar-like, first-derivative criterion. Carrying it over to the gravitational field requires a reconstruction of the metric perturbation in a horizon-regular gauge, which we have not done. Bounded derivatives are also necessary, but not sufficient, for the backreaction to be small. Other criteria give other thresholds (Table~\ref{tab:regularity}): a bounded second derivative, which stands in for the tidal curvature, requires $\omega_I\geq2\kappa$ for a generic power, while a finite energy flux integrated across the horizon requires only $\omega_I>\kappa/2$. The near-horizon form of the field is

\begin{eqnarray}
\psi (x,t) = A_{\text{in}} e^{-i \omega (t + x)} + A_{\text{out}} e^{-i \omega (t-x)} .
\end{eqnarray}

Using the coordinate transform from Rindler to Minkowski,
\begin{eqnarray}
T = \kappa^{-1} e^{\kappa x} \sinh{\kappa t}, \ \ X = \kappa^{-1} e^{\kappa x} \cosh{\kappa t},
\end{eqnarray}

we get
\begin{eqnarray}
\!\!\!\!\!\!\!\!\!\!\!\! \psi (x,t) = A_{\text{in}} [\kappa (X+T)]^{-i\omega/ \kappa} + A_{\text{out}} [ \kappa (X - T) ]^{i \omega/ \kappa}.\nonumber \\ \label{eq03}
\end{eqnarray}

A decaying mode has $\omega = \omega_R - i \omega_I$ with $\omega_I >0$. The backreaction of the perturbation on the metric depends on the energy-momentum tensor, which is built from first derivatives of $\psi$ with respect to the Minkowski coordinates $X$ and $T$. Near the future horizon, $X=T$, the outgoing term gives
\begin{eqnarray}
| \partial \psi| \propto |A_{\rm out}|\, |X-T|^{\omega_I/\kappa-1} .
\end{eqnarray}

For $\omega_I<\kappa=2\pi T_{\rm H}$ the derivative, and with it the energy density seen by an infalling observer, diverges at the horizon unless $A_{\rm out}=0$. For $\omega_I>\kappa$ the derivative is bounded and vanishes at the horizon, and regularity imposes nothing. In the marginal case $\omega_I=\kappa$ with $\omega_R\neq0$ the derivative is bounded but oscillates as $|X-T|^{i\omega_R/\kappa}$, with no limit at the horizon. Table~\ref{tab:regularity} lists the regularity classes of a single outgoing power, written in terms of the null coordinate $U\equiv T-X$, which vanishes on the future horizon.

\begin{figure*}[t]
  \centering
  \includegraphics[width=\textwidth]{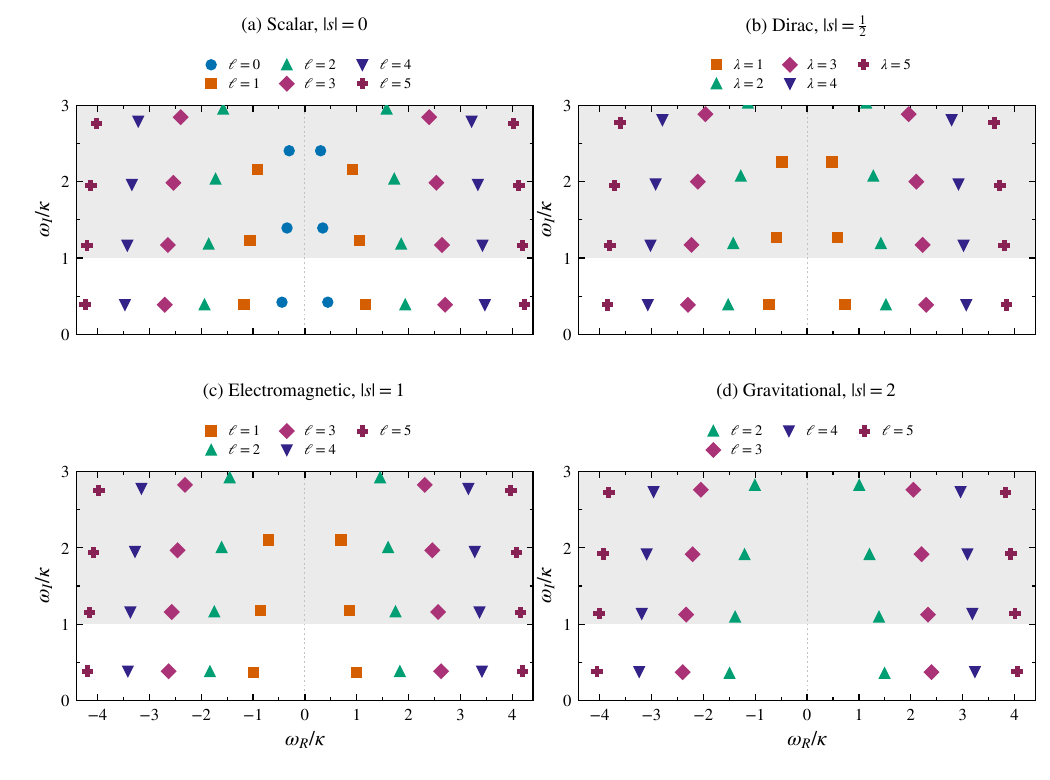}
  \caption{Schwarzschild quasinormal-mode spectra for fields of spin
    $|s|=0,\tfrac12,1,2$, with $\omega=\omega_R-i\omega_I$ and
    $\kappa=1/(4M)$ in units $G=c=1$. The angular ranges are $\ell=0,\ldots,5$ for the scalar field,
    $\lambda=1,\ldots,5$ for the Dirac field, $\ell=1,\ldots,5$ for the
    electromagnetic field, and $\ell=2,\ldots,5$ for gravitational metric
    perturbations. For the Dirac field, $\lambda$ is the magnitude of
    the angular separation constant. Colors and symbols distinguish
    angular indices. All panels use the same vertical range,
    $0\leq\omega_I/\kappa\leq3$; only modes within this window are shown.
    For each angular index, the overtone number increases from bottom
    to top, starting with $n=0$. Negative-real-frequency points are
    the mirror modes $\omega\mapsto-\bar\omega$. The gray region marks $\omega_I/\kappa\geq1$, where the scalar-like
    bounded-first-derivative criterion discussed in the text does not
    exclude an outgoing component. This shading does not change the
    retarded QNM boundary conditions. All displayed fundamentals lie
    below this boundary and all displayed overtones above it.
    Integer-spin frequencies were evaluated with \texttt{qnm}~\cite{Stein2019qnm};
    the Dirac frequencies are taken from Ref.~\cite{Jing:2005dt},
    converted to the same normalization. Some symbols near the boundary
    are closely spaced; no points have been displaced.}
  \label{fig:schwarzschild_fields}
\end{figure*}

The scope of this result is limited, and it helps to state it plainly. It concerns a single separated outgoing solution and its behaviour across the future horizon, at a generic point of the horizon away from the bifurcation surface. It says nothing about the retarded Green's function, whose poles---all the QNMs, overtones included---are fixed by causal boundary conditions. The threshold $\omega_I=\kappa$ is a boundary of local admissibility drawn on top of the standard QNM spectrum (Fig.~\ref{fig:schwarzschild_fields}). Below it, the outgoing component of any admissible solution must vanish. Above it, outgoing solutions with bounded derivatives exist alongside the QNMs, and whether they are present is decided by the dynamics, not by regularity (Sec.~\ref{sec:forced}). That the set of QNMs appearing as isolated eigenvalues depends on the regularity of the function space is known from explicit examples~\cite{Warnick:2013hr,Gajic:2019oem}; we make no spectral claim of that kind here.

For a Schwarzschild black hole only the fundamental QNM ($n=0$) of each multipole lies below the boundary (Fig.~\ref{fig:schwarzschild_fields}). The WKB approximation shows why this continues to hold as the angular momentum number $\ell$ grows: the damping rate tends to a constant, $\omega_I\to(n+\tfrac12)\lambda_L$, where $\lambda_L=1/(3\sqrt3M)$ is the Lyapunov exponent of the light ring, so that $\omega_I/\kappa\to0.77\,(n+\tfrac12)$. In this limit exactly one mode per multipole, the fundamental, lies below $\kappa$. Figure~\ref{fig:schwarzschild_fields} shows the same separation for the displayed angular indices of the scalar, Dirac, electromagnetic, and gravitational fields.
\begin{table}[t]
\caption{Regularity of a single outgoing power $\psi_{\rm out}\propto(-\kappa U)^{p}$ at a generic point of the future horizon, with $p=\alpha+i\beta$, $\alpha=\omega_I/\kappa$ and $\beta=(\omega_R-m\Omega_{\rm H})/\kappa$ (for Schwarzschild, $\Omega_{\rm H}=0$). The monomials $p=0$ and $p=1$ are exceptions to the generic conditions, as indicated. The criterion used in this paper is the fourth row.}
\label{tab:regularity}
\begin{ruledtabular}
\begin{tabular}{ll}
condition & requirement \\
\hline
bounded field & $\alpha\ge0$ \\
continuous field & $\alpha>0$, or $p=0$ \\
square-integrable $\partial_U\psi$ (integrated flux) & $\alpha>\tfrac12$, or $p=0$ \\
bounded $\partial_U\psi$ (energy density) & $\alpha\ge1$, or $p=0$ \\
continuous $\partial_U\psi$ ($\mathcal{C}^1$) & $\alpha>1$, or $p=0,1$ \\
bounded $\partial_U^2\psi$ (tidal proxy) & $\alpha\ge2$, or $p=0,1$ \\
smooth ($\mathcal{C}^\infty$) & $p=0,1,2,\dots$ \\
\end{tabular}
\end{ruledtabular}
\end{table}

When a black hole is perturbed, by shooting in a particle or by merging with another black hole, the energy needed to excite an outgoing component with $\omega_{I} < \kappa$ grows without bound toward the horizon. Such components are therefore not produced, and below the threshold only the QNMs remain. For $\omega_{I}\geq\kappa$ the derivatives are finite and regularity permits an outgoing component; whether it is present depends on the source. Hyperbolic evolution preserves regularity, so an outgoing power with non-integer $\omega_I/\kappa$, which is not smooth in $U$ across the horizon, can only come from data that are not smooth there. A smooth source can only produce the smooth powers, which we identify next.

After a binary black hole merges, the remnant settles down (in the ringdown stage) to its stationary form, and the distortion of the space-time outside it is radiated as damped modes, which have an ingoing and an outgoing part near the horizon. The ingoing part, and the nonlinear terms it generates (the stress tensor of the gravitational waves, backreacting on the metric), perturb the near-horizon geometry as they approach the horizon. This is the part of the ringdown that concerns us here, and it is not described by QNMs. At second order, the stress of the ingoing first-order field is a smooth source falling through the horizon. It generates outgoing components of the second-order field near the horizon, which Sec.~\ref{sec:forced} identifies with the horizon modes of Sec.~\ref{Horizon Modes}. It acts on the second-order field, not on the first-order ringdown.

The constraint above applies to a single separated mode. For a superposition the energy density must be evaluated for the full real field, schematically
\begin{eqnarray}
T_{\mu\nu } \sim \sum_n (\partial \psi_n)^2 + \sum_{m\neq n} \partial \psi_m \partial \psi_n .
\end{eqnarray}
It is a positive quadratic form in the first derivatives of the total field, so a divergence can be avoided only if the divergent contributions cancel in the total derivative, and an infinite sum of modes need not converge uniformly. We do not pursue this here.

\subsection{\label{Horizon Modes}Horizon Modes}

In the previous part we required the backreaction stress tensor to be finite. Here we ask for more, namely that the solution be smooth ($\mathcal{C}^{\infty}$) across the horizon, and see how this restricts it. Berti et al. \cite{Berti:2009kk} showed how a smoothness condition restricts the outgoing modes and gives the standard QNM boundary condition, but the point has received little attention since. We find that the outgoing part of Eq.~(\ref{eq03}) is smooth only for $\omega_{R}=0$ and $\omega_{I}=j\kappa$ (where $j \in \mathbb{N}$). In other words, alongside the standard QNM spectrum there is a ladder of horizon modes with frequencies $\omega_j=-ij\kappa$. In terms of the Kruskal coordinate $U=-\kappa^{-1}e^{-\kappa(t-x)}$ they are the outgoing monomials $(-\kappa U)^{j}$.

Smoothness should not be confused with analyticity, and neither turns a monomial into a propagating mode. Smoothness of a single outgoing power selects the monomials, and this is a local statement. A smooth function of $U$ has a Taylor expansion in the monomials to any finite order $N$, with a remainder $o(U^N)$ that is smaller than the last term kept. The expansion converges only if the function is analytic, which a smooth function need not be: $\exp(-1/U^2)$ is smooth and has a vanishing Taylor series. Finally, whether a monomial at the horizon connects to an outgoing wave at infinity is a global question, which we take up in Sec.~\ref{sec:forced}. We leave the $j=0$ term out of the ladder because it does not decay: it is a perturbation that co-rotates with the horizon generators, $\omega=m\Omega_{\rm H}$, and belongs to the stationary state, not to the transient. Further, note that the ladder has the spacing of the bosonic Matsubara frequencies at the Hawking temperature $T_{\rm H}=\kappa/2\pi$. However, we attach no thermal interpretation to this and do not extend it to half-integer-spin fields.

The ladder frequencies are also special for the radial equation, and this can be checked directly. Writing the perturbation as $\psi=e^{-i\omega v}R(r)$, where $v=t+r_*$ is the advanced time, which is regular across the future horizon, the Regge--Wheeler and Zerilli equations become $fR''+(f'-2i\omega)R'-(V/f)R=0$, with $f=1-2M/r$. Near the horizon we expand $R=x^{\nu}(a_0+a_1x+a_2x^2+\dots)$ in $x=r-2M$. The equation allows two leading powers, $\nu=0$ for the ingoing solution and $\nu=i\omega/\kappa$ for the outgoing one, and it fixes each coefficient $a_n$ in terms of the earlier ones. At a ladder frequency $\omega_j=-ij\kappa$ the outgoing power is the integer $\nu=j$. The outgoing solution is then the monomial $x^j$ times a power series, and it always exists. The ingoing series, however, runs into the outgoing one at order $j$, where the equation for $a_j$ reads $0\cdot a_j=c_j$, with $c_j$ fixed by $a_0,\dots,a_{j-1}$. Unless $c_j$ vanishes, no power series exists, and the ingoing solution contains a term $x^j\ln x$, which is not smooth at the horizon. We have computed $c_j$ for $\ell=2,3$ and $j=1,2,3$, for both equations, and it is nonzero in all twelve cases; we have not examined higher rungs. At these frequencies the usual roles are therefore reversed: the outgoing solution is the smooth one, and the ingoing solution, which normally defines regularity at the horizon, is not. This is the Schwarzschild counterpart of the poles of the ingoing Kerr solution and of its connection coefficients at the Matsubara frequencies~\cite{Motohashi2026KerrPoles}, and it is why the ladder frequencies are exceptional for the scattering problem: the ingoing solution cannot be normalized there in the usual way. Which of the two solutions is present near the horizon of a settling remnant is the subject of Sec.~\ref{sec:forced}.

\begin{figure}
\includegraphics[width=0.47\textwidth]{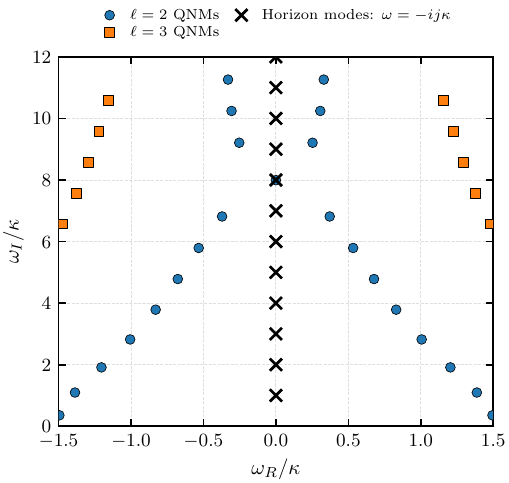}
\caption{Ladder frequencies $\omega=-ij\kappa$ (crosses on the imaginary axis) together with the QNMs~\cite{Leaver1985} of the $\ell=2$ and $\ell=3$ metric perturbations of a Schwarzschild black hole. The algebraically special frequency of $\ell=2$, $-8i\kappa$, lies on the ladder, and the QNM branch approaches it. That of $\ell=3$, $-40i\kappa$, is outside the plotted range.}
\label{HorizonModes}
\end{figure}

Fig.~\ref{HorizonModes} shows the standard QNMs of a Schwarzschild black hole together with the horizon modes. The QNM branches approach the imaginary axis near the ``algebraically special'' frequencies, which separate the lower branch of QNMs from the upper one \cite{10.2307/2397739} and are located at $\omega/\kappa=-i(\ell-1)\ell(\ell+1)(\ell+2)/3$ \cite{Cook:2016fge,Cook:2016ngj,Berti:2014bla}. These are integer multiples of $\kappa$ for every $\ell$ ($-8i\kappa$ for $\ell=2$, $-40i\kappa$ for $\ell=3$), so they fall on the ladder, but the coincidence is arithmetical. The algebraically special frequency is defined by a property of the differential equation, the vanishing of the Teukolsky--Starobinsky constant. The Regge--Wheeler and Zerilli equations behave differently there~\cite{MaassenvandenBrink:2000iwh}, and whether it is a QNM is not settled by its lying on the ladder.

\subsection{Kerr black holes}
\label{sec:forced_kerr}
For a Kerr black hole with horizon angular velocity $\Omega_{\rm H}$, the chart that is regular across the future horizon is that of ingoing Kerr coordinates $(v,r,\theta,\tilde\phi)$, with $dv=dt+(r^2+a^2)\,dr/\Delta$ and $d\tilde\phi=d\phi+a\,dr/\Delta$. A separated perturbation $\propto e^{-i\omega t+im\phi}$ becomes $e^{-i\omega v+im\tilde\phi}$ times a radial factor, and near the horizon, where the generators rotate at $\Omega_{\rm H}$, it oscillates at the co-rotating frequency $\varpi\equiv\omega-m\Omega_{\rm H}$. We exclude the extremal limit $\kappa\to0$, in which the Rindler approximation of Sec.~\ref{Backreaction limit on quasinormal modes} fails. The outgoing solution then behaves as
\be
\psi_{\rm out}\propto e^{-i\varpi(t-r_*)}\propto(-\kappa U)^{i\varpi/\kappa},\quad U\equiv-\kappa^{-1}e^{-\kappa(t-r_*)},
\label{eq:kerr_out}
\ee
where $U$ is the Kruskal-type coordinate that is regular across the future horizon, $U=0$, and a regular angular factor has been suppressed. Since ${\rm Im}\,\varpi={\rm Im}\,\omega=-\omega_I$, the argument of Sec.~\ref{Backreaction limit on quasinormal modes} carries over unchanged: $|\partial_U\psi_{\rm out}|\propto|U|^{\omega_I/\kappa-1}$, so the purely ingoing condition is enforced only for $\omega_I<\kappa$. (Here $\psi$ denotes a field that is itself regular at the horizon, such as a metric perturbation in a horizon-regular gauge; for the Teukolsky variable of spin weight $s$ the near-horizon exponents are shifted by $s$.) Likewise, smoothness at $U=0$ requires $i\varpi/\kappa\in\mathbb{N}$, i.e.,
\be
\omega^{\rm H}_j=m\Omega_{\rm H}-ij\kappa=m\Omega_{\rm H}-2\pi i\,j\,T_{\rm H},\qquad j\in\mathbb{N},
\label{eq:matsubara}
\ee
which reduces to the horizon modes $\omega=-ij\kappa$ of Sec.~\ref{Horizon Modes} for a Schwarzschild black hole ($\Omega_{\rm H}=0$). These are the Matsubara frequencies of the Hawking temperature $T_{\rm H}=\kappa/2\pi$. This is a ladder of local frequencies, not by itself a spectrum of modes radiated to infinity. The same frequencies have appeared, as ``horizon'' or ``redshift'' modes, in perturbative studies of particles plunging into Schwarzschild and Kerr black holes~\cite{Mino2008Plunge,Zimmerman2011HorizonMode,Oshita2025DirectWaves,DeAmicis2026DynQNM,Kuntz2026RedshiftModes,Kubota2026HorizonCancel}; the $j=1$ mode found in Ref.~\cite{Mino2008Plunge} was later shown to be cancelled by a zero of the source amplitude~\cite{Zimmerman2011HorizonMode}. At these frequencies one of the two Frobenius series of the radial equation at the horizon becomes resonant, so the homogeneous solutions and connection coefficients acquire poles there---for the $s=-2$ Teukolsky variable starting at $j=1-s=3$---which cancel in the Green's function~\cite{Motohashi2026KerrPoles}. In plunge waveforms the corresponding source poles are met by zeros of the source prefactor or of the response, so the waveform has no spectral line at these frequencies~\cite{Weller2026DrivenToFree,Kubota2026HorizonCancel}.

\begin{figure}[t]
  \includegraphics[width=\columnwidth]{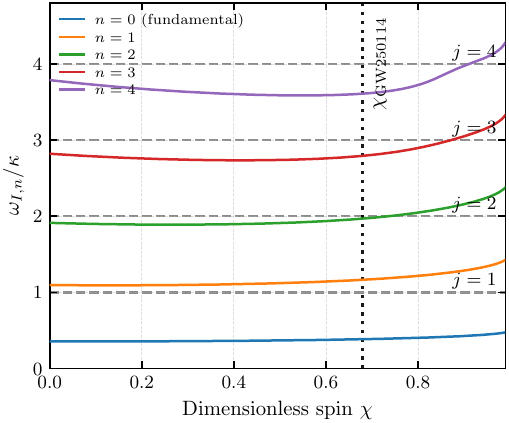}
  \caption{%
    Damping rates of the prograde $(2,2,n)$ Kerr QNMs in units of the
    surface gravity, $\omega_{I,n}/\kappa$, as a function of the
    dimensionless spin $\chi$, for overtones $n=0,\dots,4$. Horizontal
    dashed lines mark the horizon (Matsubara) ladder $\omega_I = j\kappa$
    with $j=1,2,3,4$. The vertical dotted line indicates the remnant spin
    of GW250114, $\chi = 0.68$. For every spin, only the fundamental
    mode ($n=0$) lies below the regularity threshold $\omega_I/\kappa =
    1$, while all overtones lie above it. This is the Kerr counterpart
    of Fig.~\ref{fig:schwarzschild_fields}. The frequencies were computed using Leaver's continued-fraction method as implemented in \texttt{qnm}~\cite{Stein2019qnm}; at $\chi=0$, the $n=0,\dots,3$ values agree with the gravitational $\ell=2$ sequence in Fig.~\ref{fig:schwarzschild_fields}.}
  \label{fig:kerr_ladder}
\end{figure}

Figure~\ref{fig:kerr_ladder} compares the damping rates of the prograde $\ell=m=2$ Kerr QNMs with this ladder. At every spin only the fundamental mode lies below the regularity threshold: $\omega_{I,0}/\kappa$ rises from $0.36$ at $\chi=0$ to $0.48$ at $\chi=0.99$ and tends to $1/2$ at extremality, while the first overtone lies just above the threshold ($1.10$ at $\chi=0$, $1.17$ at $\chi=0.68$). The same holds for the other low-lying multipoles we have checked: $(\ell,m)=(2,1),(3,3),(3,2),(4,4)$ for $\chi\le0.99$, and $(2,0),(2,-2)$ for $\chi\le0.95$. It does not hold for every branch at every spin. For the axisymmetric and retrograde branches the damping stays finite as $\kappa\to0$~\cite{Yang2013Branching}, so $\omega_{I,0}/\kappa$ exceeds unity near extremality (at $\chi=0.99$ it is $1.18$ for $(2,0)$ and $1.42$ for $(2,-2)$), and the $(2,1)$ and $(3,2)$ branches cross the threshold between $\chi=0.99$ and $0.999$. The Schwarzschild conclusion of Sec.~\ref{Backreaction limit on quasinormal modes}, that the bounded-derivative condition enforces the ingoing boundary condition only for the fundamental mode, therefore extends to the $\ell=m=2$ branch of Kerr at all spins, and to the other branches within the spin ranges quoted. Frequencies were computed with the \texttt{qnm} package~\cite{Stein2019qnm}, with its branch labelling.

\section{Forced ringdown: driven and free contributions of a horizon-crossing source}
\label{sec:forced}

Section~\ref{sec:regularity} identified two local scales at the horizon: the damping rate $\omega_I=\kappa$, below which bounded derivatives enforce the ingoing boundary condition, and the ladder of horizon modes selected by smoothness. Both acquire a dynamical meaning once the ringdown of a merger is treated as a forced problem. In this section we set up that problem, describe how the horizon controls the source, separate the response of each QNM into a driven and a free part, and work out which modes can be observed as free oscillations and when.

\subsection{Setting and perturbative orders}
\label{sec:forced_eq}
We consider two classes of source. (i) A body of mass $\mu\ll M$ plunging into a Kerr black hole. At first order in $\mu/M$ the Teukolsky equation (or, in Schwarzschild, the Regge--Wheeler or Zerilli equation) is inhomogeneous, with a distributional source on the worldline, and the problem is under control~\cite{Mino2008Plunge,Zimmerman2011HorizonMode,Kubota2026HorizonCancel,Weller2026DrivenToFree}. (ii) A comparable-mass merger. Here there is no small parameter at the time the common horizon forms. Nor is the linear vacuum ringdown of the remnant forced by its own backreaction: in the expansion $g=g_0+\epsilon h^{(1)}+\epsilon^2h^{(2)}+\dots$ the second-order source $S^{(2)}[h^{(1)},h^{(1)}]$ acts on $h^{(2)}$ and leaves $Lh^{(1)}=0$. For class (ii) we use the particle problem as a model of the near-horizon excitation, treat its extension as a hypothesis, and test it against data in Sec.~\ref{sec:horizon}. Schematically, with the structure of the Kerr master equation suppressed,
\be
\left[-\partial_t^2+\partial_{r_*}^2-V_\ell(r)\right]\psi_{\ell m}(t,r)=S_{\ell m}(t,r),
\label{eq:forced}
\ee
and the signal at infinity is the retarded response $\tilde\psi_{\ell m}(\omega)=\tilde G_{\ell m}(\omega)\,\tilde I_{\ell m}(\omega)$. Here $\tilde G$ is the reduced response, with QNM poles $\omega_n$ of residues $g_n$, a branch cut, and a non-pole part, and $\tilde I$ is the source integrated against the ingoing radial solution~\cite{Mino2008Plunge,Zimmerman2011HorizonMode,Weller2026DrivenToFree}. In the time domain each pole contributes the causal convolution
\be
y_n(u)=-i\,g_n\int_{-\infty}^{u}e^{-i\omega_n(u-v)}\,I(v)\,dv ,
\label{eq:pole_conv}
\ee
where $u$ is retarded time at the observer, and the signal is
\be
\psi(u)=\sum_n y_n(u)+\psi_{\rm prompt}(u)+\psi_{\rm tail}(u),
\label{eq:qnm_sum}
\ee
with $\psi_{\rm tail}$ the power-law contribution of the branch cut and $\psi_{\rm prompt}$ the remaining non-pole part~\cite{Leaver1986,Andersson1997Evolving,Nollert1999Quantifying,DeAmicis2026DynQNM}. Once the source has stopped, or has decayed faster than mode $n$, the pole response becomes a free oscillation,
\be
y_n(u)\to R_n\,e^{-i\omega_n u},\qquad R_n=-i\,g_n\,\tilde I(\omega_n),
\label{eq:free_amp}
\ee
which is the constant-coefficient term of the usual QNM sum. Equations~\eqref{eq:pole_conv} and~\eqref{eq:free_amp} are different objects: the first holds at all times, the second only after the forcing of mode $n$ has ended. The free amplitude $R_n$ exists when $\tilde I(\omega)$ can be continued analytically to $\omega_n$. A source bounded by $e^{-\gamma_su}$ has a transform that is analytic for ${\rm Im}\,\omega>-\gamma_s$; for poles deeper than that the continuation has to be established separately, and the singularities of $\tilde I$ met on the way contribute to the signal. These contributions are the driven terms of Sec.~\ref{sec:forced_which}. Amplitudes must be compared at a common reference time: moving the reference time by $\delta$ multiplies $R_n$ by $e^{-i\omega_n\delta}$, which changes the convention and not the waveform.

\subsection{Horizon control of the source}
\label{sec:forced_source}
For the part of the source that crosses the future horizon, the late-time form of $I(u)$ is fixed by the horizon. Along an ingoing geodesic $r-r_+\propto e^{-2\kappa t}$ in Boyer--Lindquist time, which is $e^{-\kappa u}$ in retarded time at infinity, so the position of the source is an analytic function of the Kruskal coordinate $U\propto-e^{-\kappa u}$ (cf.~\cite{Mino2008Plunge}). Analyticity of the trajectory does not by itself make the projected source analytic in $U$. For a plunging particle, however, the perturbative treatments find that $\tilde I(\omega)$ has a tower of poles at the Matsubara frequencies~\cite{Mino2008Plunge,Zimmerman2011HorizonMode,Kubota2026HorizonCancel}, which in the time domain is a sum over the ladder,
\be
I(u)\;\simeq\;\sum_{j\ge j_0}b_j\,e^{-i\omega^{\rm H}_j u}\qquad(u>u_c).
\label{eq:source_ladder}
\ee
The first rung present, $j_0$, depends on the field and on how the problem is split into source and response, as we discuss below. This does not conflict with Sec.~\ref{sec:regularity}: the ladder there is a local statement about a field that is regular at the horizon, which holds without linearization and starts at $j=1$, whereas $j_0$ records which rungs drive radiation to infinity in linear theory. The time $u_c$ beyond which the expansion is useful is set by the singularities of the projected source in the $U$-plane. Sources outside the horizon are not bound to the ladder. Quadratic couplings of exterior QNMs radiate at sum and difference frequencies, with their own angular selection rules (the product of two $(2,2,0)$ modes has $m=4$), and the late-time tails decay as power laws. These are distinct components, outside the forcing considered here.

What the horizon fixes is the rate. The time at which the source is within a relative tolerance of its leading asymptotic form is
\be
u_\mathrm{tol}-u_{\rm ref}=\kappa^{-1}\left[\ln(1/\mathrm{tol})+c+O(\mathrm{tol})\right],
\label{eq:utol}
\ee
with a constant $c$ that depends on the trajectory. For geodesic plunges into a Schwarzschild black hole we take $u_{\rm ref}$ at the light-ring crossing and define the tolerance as the relative deviation of $(r-2M)e^{\kappa u}$ from its limit. We find $c=0.18$, $0.89$, $1.47$ and $1.62$ for a radial plunge from rest at infinity, a plunge from infinity with $L=3M$, a plunge from just inside the innermost stable circular orbit (ISCO; $E=\sqrt{8/9}$, $L=2\sqrt3M$), and a near-critical plunge with $L=3.9M$. A tolerance of $10\%$ is reached $2.3$--$3.8/\kappa$ after the light-ring crossing, and $1\%$ after $4.8$--$6.2/\kappa$ (Table~\ref{tab:plunge}). The horizon therefore sets the scale $1/\kappa$, with a logarithmic prefactor that depends on the tolerance. It does not set a universal duration, and the time at which the waveform starts to ring freely has to be calibrated separately (Sec.~\ref{sec:forced_obs}).

\begin{table}[t]
\caption{Retarded time after the light-ring crossing at which $(r-2M)e^{\kappa u}$ along a Schwarzschild geodesic plunge is within the stated relative tolerance of its limit, in units of $1/\kappa=4M$. For small tolerances the entries follow Eq.~\eqref{eq:utol}; at the largest ones the $O(\mathrm{tol})$ correction is visible.}
\label{tab:plunge}
\begin{ruledtabular}
\begin{tabular}{lcccc}
plunge & tol $0.3$ & $0.1$ & $0.03$ & $0.01$\\
\hline
radial, from rest at $\infty$ & 0.87 & 2.33 & 3.64 & 4.77\\
from infinity, $L=3M$ & 1.61 & 3.05 & 4.36 & 5.48\\
from the ISCO & 2.23 & 3.64 & 4.94 & 6.06\\
near-critical, $L=3.9M$ & 2.38 & 3.79 & 5.08 & 6.21\\
\end{tabular}
\end{ruledtabular}
\end{table}

In the particle problem the Matsubara poles of $\tilde I(\omega)$ do not survive in the signal at infinity. In Kerr with spin weight $s=-2$, the $j=1,2$ poles are cancelled by zeros of the source prefactor and the $j\ge3$ poles by zeros of the inverse connection coefficient $1/B^{\rm inc}$, whatever the orbit and the black hole parameters~\cite{Zimmerman2011HorizonMode,Kubota2026HorizonCancel,Motohashi2026KerrPoles}. In Schwarzschild, at the level of the metric perturbation, the cancellation follows from causality, which forces the source-integrated Green function to vanish on the light cone; the individual overtones carry horizon-frequency pieces that cancel only in their sum together with the impulsive term~\cite{Kuntz2026RedshiftModes}. Two features of these results matter below. First, a zero that cancels a pole leaves a finite contribution, not a vanishing one. What vanishes is the residue at the horizon frequency, that is, the stationary spectral line. The finite-time information about the trajectory stays in the waveform. Second, the split of the waveform into source and response is not unique: frequency-dependent factors can be moved from one to the other, which changes the apparent decay rate of the source without changing the waveform. For a scalar field the ladder starts at $j_0=1$. For gravitational waves the factors $(\omega-\omega^{\rm H}_1)(\omega-\omega^{\rm H}_2)$ belong to the source in the Teukolsky description and to the transmission through the potential barrier in the Sasaki--Nakamura description~\cite{Nakamoto2026PoleSplitting}, and in either description the first rung present in the source integral that drives the QNM poles is $j_0=3$~\cite{Weller2026DrivenToFree,Laeuger2026SourceImprints}. We use these results within their stated scope and do not extend them to the second-order sources of comparable-mass mergers.

The ladder describes the source once it is close to the horizon. Earlier in the plunge the source drives the exterior at an instantaneous frequency that follows its motion. A saddle-point evaluation of the source integral gives~\cite{Oshita2025DirectWaves,Sun2026DWComplexFreq,Nakamoto2026PoleSplitting,Laeuger2026SourceImprints}
\be
\omega_{\rm G}(u)=m\,\frac{\Omega+\beta\,\Omega_{\rm H}}{1+\beta}-2i\kappa\,\frac{\beta}{1+\beta},
\label{eq:omegaG}
\ee
where $\Omega=d\phi/dt$ is the angular velocity of the source and $\beta=-dr_*/dt$ its inward velocity in the tortoise coordinate. Since $0<\beta<1$, the decay rate $2\kappa\beta/(1+\beta)$ grows steadily toward $\kappa$ and never exceeds it, and $\omega_{\rm G}\to m\Omega_{\rm H}-i\kappa$, the first horizon mode, as the source reaches the horizon. Equation~\eqref{eq:omegaG} describes the forcing while the source is outside the potential barrier. Once the source is inside, the barrier screens the first two rungs and the forcing of the poles decays at $3\kappa$~\cite{Nakamoto2026PoleSplitting,Laeuger2026SourceImprints}. The changeover takes a few $M$: for prograde equatorial plunges from the ISCO at $\chi=0.5$--$0.9$, the complex frequency of the source integral evolves rapidly over about $10M$ around the light-ring crossing, and the late emission lags the light cone by about $3M$~\cite{Laeuger2026SourceImprints}. Equation~\eqref{eq:omegaG} is a leading-order description, and the saddle-point approximation behind it fails very close to the horizon~\cite{Kubota2026HorizonCancel}; a treatment that keeps the full frequency dependence of the source is given in Ref.~\cite{Laeuger2026SourceImprints}. It is sufficient for the estimates that follow.

\subsection{Driven and free contributions}
\label{sec:forced_which}

Differentiating Eq.~\eqref{eq:pole_conv}, the response of each pole obeys
\be
\dot y_n+i\omega_n y_n=-i g_n I(u),\qquad \omega_n=\Omega_n-i\gamma_n .
\label{eq:pole_ode}
\ee
For a drive $I(u)=I_0e^{-i\omega_s(u-u_0)}$ switched on at $u_0$, with $\omega_s=\Omega_s-i\gamma_s\neq\omega_n$ and $y_n(u_0)=0$, direct integration gives, for $u\ge u_0$,
\be
y_n(u)=\frac{g_n I_0}{\omega_s-\omega_n}\left[e^{-i\omega_s(u-u_0)}-e^{-i\omega_n(u-u_0)}\right],
\label{eq:driven_free}
\ee
a driven term at the source frequency and a free term at the QNM frequency, with
\be
\frac{|y_{n,\rm driven}|}{|y_{n,\rm free}|}=e^{-(\gamma_s-\gamma_n)(u-u_0)} .
\label{eq:ratio}
\ee
The free term wins at late times if and only if $\gamma_s>\gamma_n$. If the pole was already excited at $u_0$, or the drive has several terms, the driven and free parts carry coefficients $D_n$ and $C_n$, and the ratio becomes $|D_n/C_n|\,e^{-(\gamma_s-\gamma_n)(u-u_0)}$. For $C_n\neq0$ the free part then exceeds the driven part by a factor $1/\varepsilon$ after
\be
\Delta u\simeq\frac{\ln\left(|D_n/C_n|/\varepsilon\right)}{\gamma_s-\gamma_n}.
\label{eq:sep_time}
\ee
At exact resonance, $\omega_s=\omega_n$, the response is $y_n=-ig_nI_0(u-u_0)e^{-i\omega_n(u-u_0)}$. Equal damping with different real parts gives instead two exponentials with the same envelope, which beat. For a source that is a sum of exponentials, $I(u)=\sum_j b_je^{-i\omega_j u}\theta(u-u_0)$, the driven terms add up to $\sum_j b_j\,\tilde G_{\rm pole}(\omega_j)\,e^{-i\omega_j u}$ with $\tilde G_{\rm pole}(\omega)=\sum_n g_n/(\omega-\omega_n)$, plus onset terms at the QNM frequencies. Three consequences follow.
\begin{enumerate}
\item[(i)] The drive is visible in the waveform only where the response has no zero. Summing the driven terms of all poles, together with the non-pole part of the response, gives the source amplitude times the full response at the source frequency. For the plunging sources of Sec.~\ref{sec:forced_source} this product vanishes at the ladder frequencies, so once the source is on the ladder the waveform is a QNM sum with the constant amplitudes $R_n$ of Eq.~\eqref{eq:free_amp}. It is then indistinguishable from free ringing, even though each pole is still driven in the sense of Eq.~\eqref{eq:pole_ode}~\cite{Weller2026DrivenToFree}. Before screening the source drives at $\omega_{\rm G}(u)$, where the response has no zero, and the driven part appears in the waveform.
\item[(ii)] The fundamental mode separates from a horizon-controlled source early. For the $\ell=m=2$ branch $\gamma_0<\kappa/2$ at all spins (Fig.~\ref{fig:kerr_ladder}). The drive of Eq.~\eqref{eq:omegaG} decays faster than the fundamental mode once $2\beta/(1+\beta)>\gamma_0/\kappa$, which at $\chi=0.68$ means $\beta>0.24$, and every rung of the ladder decays faster still. Whichever rung leads in a given description of the source, the free part of the fundamental mode takes over, after the time given by Eq.~\eqref{eq:sep_time}. More generally, any mode with $\gamma_n<\kappa$ and $C_n\neq0$ separates from any source whose projection decays at $\gamma_s\ge\kappa$. This is the dynamical content of the regularity threshold of Sec.~\ref{Backreaction limit on quasinormal modes}.
\item[(iii)] An overtone has $\gamma_n>\kappa$, so it decays faster than the drive of Eq.~\eqref{eq:omegaG} at every stage of the plunge. Until the source is screened its response is dominated by the driven part, which by (i) appears in the waveform. After screening the source is on the ladder, the driven terms cancel, and the waveform is a QNM sum with constant amplitudes. What can be observed of an overtone is the amplitude it has left by then. For the GW250114 remnant ($\chi=0.68$, $1/\kappa=4.73M_f$) a free mode retains, over an interval $1/\kappa$, the fractions $e^{-\gamma_n/\kappa}=0.68$, $0.31$, $0.14$ and $0.06$ of its amplitude for $n=0,1,2,3$. These are amplitude ratios, not detection probabilities. Whether what remains can be measured depends on the excitation, the signal-to-noise ratio, the covariance with other components and the template error, and a fraction of $14\%$ is not a prohibition.
\end{enumerate}
Three statements about an overtone are often run together, and we keep them apart. An overtone may be (a) still driven at the start time of a fit; (b) not separately identifiable in a given data set; (c) a resonance of the black hole regardless, as every QNM is. We claim (a) for fits that start before the source is screened. Whether (b) holds has to be decided case by case by the criteria just listed, and nothing here argues against (c). Related results point the same way. The collective behaviour of the overtones changes when a plunging source crosses $r_*=0$: before the crossing successive overtones are driven in counter-phase and interfere destructively, and their amplitudes remain time dependent for some time after it~\cite{DeAmicis2026DynQNMII}. Quadratic modes can exceed the linear overtones~\cite{Cheung2023Nonlinear,Mitman2023Nonlinear,Bourg2026Dissection}. And in the models studied in Ref.~\cite{Nee2023OvertoneRole}, additional damped sinusoids absorb non-QNM structure even when their frequencies are not those of the system. On the other hand, accurate NR waveforms give overtone amplitudes that are stable over a range of start times~\cite{Giesler2025OvertonesNonlin,Dyer2025ModeContent}. Our argument does not contradict this: by (i) and (iii) the waveform is a QNM sum with constant amplitudes once the source is screened.

\subsection{The filtered residual and the ``direct wave''}
\label{sec:forced_dw}

The driven part of the signal can be exposed by filtering QNMs out of the waveform. The rational filter $F(\omega)=\prod_n(\omega-\omega_n)/(\omega-\bar\omega_n)$ has zeros at the chosen QNM frequencies and unit modulus on the real axis~\cite{Ma2022QNMFilters,Ma2023ModeCleaning}. A single factor has the time-domain kernel $\delta(t)-2\gamma_ne^{-i\bar\omega_nt}\Theta(-t)$. It is an all-pass filter, not a projection: it moves signal to earlier times and rescales the poles it does not remove by $F(\omega_k)$, which need not have unit modulus. The instantaneous complex frequency of the residual is a ratio of interfering sums, and it can rise, fall or diverge near a node. The residual left after removing the $(2,2,0)$, $(2,2,1)$ and $(2,2,2)$ modes from NR waveforms is what has been called the ``direct wave''. It has been interpreted as a signal from the plunge near the horizon, with the instantaneous frequency of Eq.~\eqref{eq:omegaG}~\cite{Oshita2025DirectWaves}. In NR waveforms its complex frequency is reported to evolve toward $2\Omega_{\rm H}$ in real part and toward about $2\kappa$ in damping, reaching about $3\kappa$ in some high-spin cases~\cite{Sun2026DWComplexFreq}. An analysis in Schwarzschild traces part of it to the anti-causal poles of the filter, while arguing that the rest carries information about the source near the horizon~\cite{MaWang2026FoundationsDW}. It has been reported in GW250114 as a signature of the post-merger horizon~\cite{Lu2026GW250114Horizon} and used to measure the area of the remnant~\cite{Chung2026DWAreaLaw}.

Whether the direct wave measures properties of the horizon is an open question, and several results call for caution. The exact cancellation results of Sec.~\ref{sec:forced_source} show that no stationary line at a horizon frequency survives at infinity, in Schwarzschild~\cite{Kuntz2026RedshiftModes} or in Kerr~\cite{Kubota2026HorizonCancel}; they do not show that the residual is free of information about the source. In NR waveforms the fitted frequency and damping of the residual follow $\Omega_{\rm H}$ and $\kappa$ only near $\chi_f\approx0.7$~\cite{Kankani2026NotMeaningful}, which happens to be the remnant spin of GW250114. They drift by $10$--$20\%$ with the fit window, although a horizon-frequency template remains useful in practice~\cite{Dyer2026ModelingDW}. The damping measured over a finite window can lie below $\kappa$~\cite{Han2026RedshiftTransfer}, as Eq.~\eqref{eq:omegaG} leads one to expect. After the $n=0,1,2$ modes have been removed, the residual is also hard to tell apart from the next overtone: $\omega_{I,3}=(2.7$--$3.3)\,\kappa$ over $0\leq\chi\leq0.99$ in Fig.~\ref{fig:kerr_ladder}, and the real part of $\omega_{223}$ lies within $3\%$ of $2\Omega_{\rm H}$ for $\chi\gtrsim0.9$ and within $17\%$ at the spin of GW250114~\cite{Weller2026DrivenToFree}. How well the two can be separated depends on the window and on the signal-to-noise ratio, and has not been quantified.

The most recent analyses bring the direct-wave literature close to the picture of this section. Refs.~\cite{Nakamoto2026PoleSplitting,Laeuger2026SourceImprints} describe the direct wave as the forced response at the instantaneous frequency of the source, screened by the potential barrier, and conclude that it is not a fixed horizon mode. Ref.~\cite{Laeuger2026SourceImprints} also shows that dividing the waveform by the black hole response, instead of applying a rational filter, returns the source integral $I(u)$ itself, through a kernel that is localized on the light cone and has no anti-causal part. This is the quantity that drives the poles in Eq.~\eqref{eq:pole_ode}. In our terms the direct wave is the driven part of the signal, seen through a filter. Our argument does not depend on how the direct wave is interpreted. It rests on the convolution of Sec.~\ref{sec:forced_which} and on the cancellation and screening results within their scope.

\subsection{Predictions}
\label{sec:forced_obs}

The framework yields three testable hypotheses, stated so that each can fail.
\begin{enumerate}
\item[(a)] \textit{Onset of free ringing.} Define $u_c$ operationally as the earliest start time after which the fitted $(2,2,0)$ and $(2,2,1)$ amplitudes, referred to a common time, no longer depend on the start time within their uncertainties. We conjecture that $u_c-u_{\rm peak}=C/\kappa$, with a coefficient $C$ common to remnants of different spin. This is an empirical hypothesis and not a consequence of the geodesic scaling of Eq.~\eqref{eq:utol}: $C$ depends on the trajectory, on the tolerance, and on the delay between the light-ring crossing and the peak of the waveform. It has to be fitted, with its uncertainty, over a set of NR waveforms spanning $\chi$ (for which $1/\kappa=4.0M$, $4.7M$, $6.6M$ and $16M$ at $\chi=0$, $0.68$, $0.9$ and $0.99$) and compared with the alternatives of a fixed multiple of $M_f$, a QNM damping time, and the barrier timescale. The same onset should be visible in the waveform divided by the black hole response, whose instantaneous decay rate should rise through $\kappa$ toward $3\kappa$, as it does for test particles (Fig.~1 of Ref.~\cite{Laeuger2026SourceImprints}). In GW250114 the $(2,2,1)$ amplitude follows free decay only for start times $\gtrsim6M_f$ after the peak~\cite{LVK2026GW250114Spec}. This is consistent with such an onset, but one event near $\chi_f\approx0.7$ cannot determine the scaling.
\item[(b)] \textit{Two-component models.} Before $u_c$ the non-fundamental content of the signal is the driven response of Eq.~\eqref{eq:omegaG}. Compared with a free $(2,2,1)$ mode it oscillates more slowly and decays no faster than $\kappa$. A model made of the fundamental mode and a damped sinusoid at the horizon frequency $m\Omega_{\rm H}-i\kappa$ should then be preferred over the fundamental mode plus a free first overtone, and the preference should fade after $u_c$. Because the drive has not yet reached its horizon value, the horizon parameters returned by such a fit should differ from those of the final remnant. And because all of this happens within GR, the same behaviour must appear in injections of GR waveforms calibrated to NR, at every start time. Section~\ref{sec:horizon} carries out a first test on GW250114.
\item[(c)] \textit{Start-time dependence.} For start times before $u_c$, the fitted amplitudes of modes with $\gamma_n$ larger than the decay rate of the forcing, referred to a common time, should depend on the start time in the manner of Eq.~\eqref{eq:driven_free}. After $u_c$ they should not. This is distinct from the spectral instability of the overtones under perturbations of the operator and from the conditioning of a many-overtone fit, both of which persist after $u_c$.
\end{enumerate}

\section{Search for horizon modes in GW250114}
\label{sec:horizon}

\subsection{Analysis strategy}

We search for evidence of a $(\ell,m,j)=(2,2,1)$ horizon mode in the post-merger signal of GW250114~\cite{LVK2025GW250114} using the \texttt{PyCBC
Inference} framework~\citep{Biwer2019}. The signal is modelled hierarchically as the sum of an inspiral submodel
(\texttt{SEOBNRv5PHM}) and a ringdown submodel composed of
time-domain quasinormal modes, with a shared set of extrinsic parameters. The ringdown start time is anchored to the coalescence time $t_c$ via
\begin{equation}
    t_{\rm gate}^{\rm rd} = t_c + N \, M_f \, \frac{G M_\odot}{c^3},
    \label{eq:gate}
\end{equation}
where $M_f$ is the sampled detector-frame remnant mass in solar masses and $N$ is an integer, so that the offset is $N$ in
units of the final mass. Here $t_c$ denotes the coalescence time defined for the IMR waveforms used in this work. The excised window (the gate) has a fixed duration of $1\,$s, so the number of analysed samples and, for stationary noise, the determinant of the truncated covariance are independent of the sampled parameters; the likelihood normalisation is therefore constant and is omitted.
We perform a scan
 $N\in\{-6,-4,-2,-1,0,+1,+2,+3,+4,+6,+8\}$ to test the robustness of the
recovered remnant parameters against the assumed ringdown start
time. Full details of the data conditioning, priors, sampler
configuration, and waveform settings are given in
App.~\ref{app:method}.

Two competing ringdown hypotheses are compared, both including the
fundamental $(\ell,m,n) = (2,2,0)$ mode with free amplitude
$A_{220}$ and phase $\phi_{220}$:
\begin{itemize}
    \item an \textbf{overtone hypothesis} (220 + 221 overtone), in which the
    first overtone $(2,2,1)$ has a free amplitude $A_{221}$ and phase offset $\Delta\phi_{221}$, and the $(\ell,m,j)=(2,2,1)$ horizon mode amplitude is fixed to zero;
    \item a \textbf{horizon-mode hypothesis} (220 + horizon mode, $(\ell,m,j)=(2,2,1)$), in
    which the $(2,2,1)$ horizon mode has a free amplitude $A^{\rm HM}_{221}$ and phase offset $\Delta\phi^{\rm HM}_{221}$, and the $(2,2,1)$ overtone amplitude is fixed to zero.
\end{itemize}
In both cases the remnant mass and spin are free parameters, so
the QNM frequencies are not tied to the inspiral prediction and
the ringdown sector provides a separate estimate of the
final state (the two sectors share the extrinsic parameters, so the two estimates are not statistically independent). The Bayes factor
$\mathcal{B}^{\rm HM}_{\rm OT}(N)
= \mathcal{Z}_{\rm HM}(N)/\mathcal{Z}_{\rm OT}(N)$
quantifies the relative support for the two hypotheses, with
$\mathcal{B}^{\rm HM}_{\rm OT} > 1$ favouring the horizon mode.
The inspiral-to-remnant mapping used to compare the two sectors
is described in App.~\ref{app:mapping}.

The second component of the horizon-mode hypothesis is the $j=1$ rung of the ladder of Eq.~\eqref{eq:matsubara} for $m=2$,
\be
\omega^{\rm H}=2\Omega_{\rm H}-i\kappa,
\label{eq:hyp}
\ee
evaluated, like the Kerr QNM frequencies, at the sampled $(M_f,\chi_f)$ (App.~\ref{app:ringdown_model}). It is not a relabelled overtone. At $\chi_f=0.68$ one has $M_f\kappa=0.2115$ and $2M_f\Omega_{\rm H}=0.3923$, against $M_f\omega_{220}=0.5240-0.0815i$ and $M_f\omega_{221}=0.5119-0.2465i$. At the same remnant parameters the horizon mode thus oscillates at $0.77$ of the frequency of the first overtone and decays at $0.86$ of its rate. For a detector-frame mass of $68\,M_\odot$ this is $186$\,Hz with a damping time of $1.58$\,ms, against $243$\,Hz and $1.36$\,ms for the $(2,2,1)$ mode and $249$\,Hz and $4.11$\,ms for the $(2,2,0)$ mode. The Bayes factor compares these two constrained families, each with its priors. It does not measure the frequency and damping of the second component on their own, because the remnant parameters, amplitudes and phases all adjust together. No model containing the $(2,2,2)$ mode has been run. 
For both hypotheses, we used the same uniform prior on the phase angle of the second component and the same uniform prior, over $[0,5]$, on its amplitude relative to that of the $(2,2,0)$ mode.

\subsection{Results}

\begin{figure}[t!]
    \centering
    \includegraphics[width=0.48\textwidth]{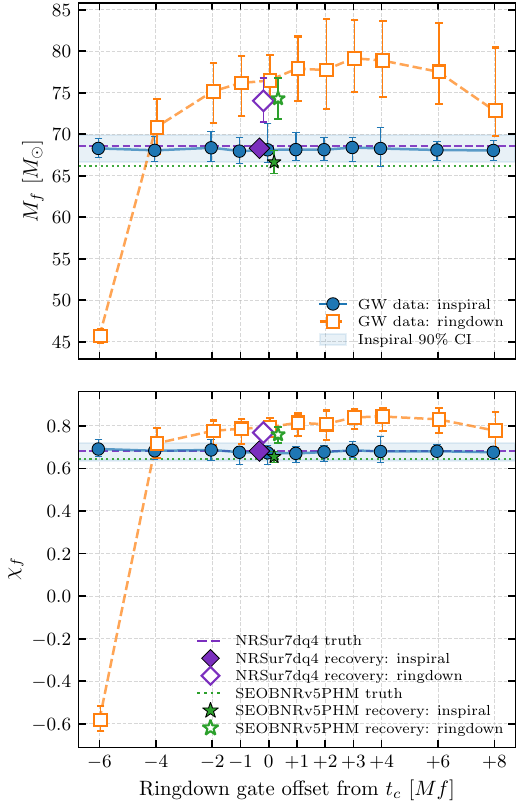}
    \caption{Final mass (top) and final spin (bottom) of the remnant black
    hole inferred from the inspiral (blue circles) and ringdown
    (orange squares) sectors of the horizon-mode analysis, as a
    function of the ringdown gate offset from $t_c$ in units of the
    final mass. Error bars are 90\% credible intervals. The shaded
    blue band marks the pooled 90\% credible interval of the
    inspiral-sector prediction across all offsets. Dashed purple and
    dotted green horizontal lines mark the injected $M_f$ and
    $\chi_f$ of the \texttt{NRSur7dq4} and \texttt{SEOBNRv5PHM}
    injections (App.~\ref{app:injection}). Their recovered values at
    $N=0$ are shown at $t_c$ as a diamond (\texttt{NRSur7dq4}) and a
    star (\texttt{SEOBNRv5PHM}), filled for the inspiral sector and
    open for the ringdown sector, offset horizontally for display.
    The displacement between an injected value and the
    corresponding recovered value measures the bias of each sector at
    zero gate offset; for the ringdown sector it is the
    template-dependent shift discussed in the text.}
    \label{fig:final_state_horizonmode}
\end{figure}

\begin{figure}[t!]
    \centering
    \includegraphics[width=0.48\textwidth]{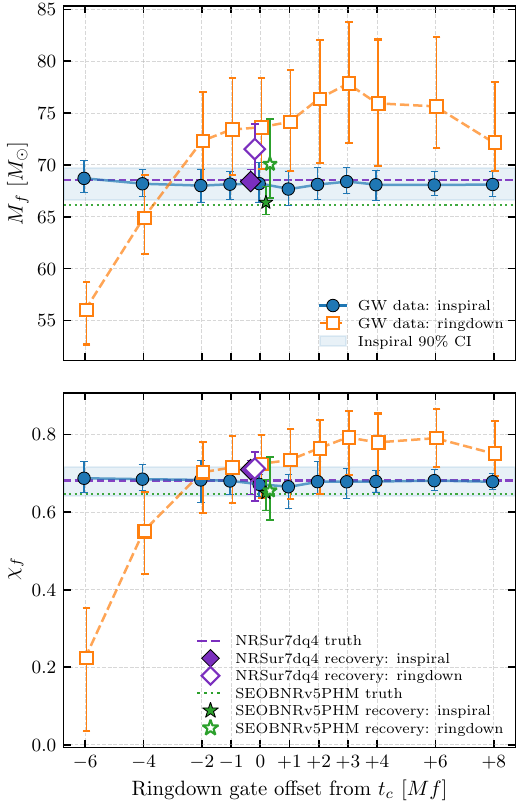}
    \caption{Same as Fig.~\ref{fig:final_state_horizonmode}, but
    for the overtone hypothesis (220 + 221).}
    \label{fig:final_state_overtone}
\end{figure}

Figure~\ref{fig:final_state_horizonmode} shows the median remnant
mass and spin inferred from the inspiral sector (circles) and the
ringdown sector (squares) as a function of the gate offset $N$ for
the horizon-mode hypothesis, with 90\% credible intervals.
Figure~\ref{fig:final_state_overtone} shows the same comparison
for the overtone hypothesis. Under the overtone hypothesis the ringdown sector returns $M_f=72.3$--$77.9\,M_\odot$ and $\chi_f=0.70$--$0.79$ for $-2\le N\le+6$, against $M_f\simeq68\,M_\odot$ and $\chi_f\simeq0.68$ from the inspiral sector. Under the horizon-mode hypothesis it returns $M_f=75.2$--$79.2\,M_\odot$ and $\chi_f=0.78$--$0.84$ over the same gates, and falls back to $M_f=72.8\,M_\odot$ (90\% interval $69.8$--$80.4$) at $N=+8$. Because the two sectors share parameters, we compare them through the paired difference between the ringdown and inspiral values, evaluated sample by sample. Under the horizon-mode hypothesis the mass difference is $+6.8$ to $+10.8\,M_\odot$ and the spin difference $+0.09$ to $+0.16$ for $-2\le N\le+6$; at $N=0$, for example, $\Delta M_f=+8.2\,M_\odot$ (90\% interval $+4.1$ to $+11.8$) and $\Delta\chi_f=+0.12$ ($+0.06$ to $+0.18$). Both differences are positive and exclude zero at the 90\% level at every gate from $N=-2$ to $+8$. Under the overtone hypothesis the mass difference is $+4.4$ to $+9.4\,M_\odot$ for $-2\le N\le+6$ and excludes zero at every gate from $N=-1$ to $+8$, whereas the spin difference, $+0.02$ to $+0.12$, excludes zero only at $N=+3$, $+6$ and $+8$. Both models fail at $N=-6$, where the ringdown segment contains the merger. Both templates thus return a heavier remnant than the inspiral predicts. The horizon mode adds a further $1$--$3\,M_\odot$ and, unlike the overtone, a spin shift that is significant from $N=-2$ on. We return to this shift below.

\begin{figure}[t!]
    \centering
    \includegraphics[width=0.48\textwidth]{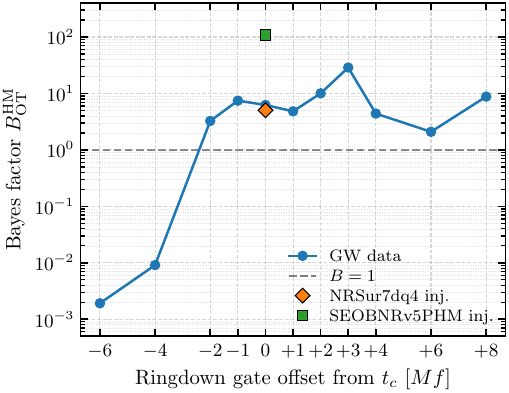}
    \caption{Bayes factor $B^{\mathrm{HM}}_{\mathrm{OT}}$ as a function of the ringdown gate offset from $t_c$. The dashed line marks $B=1$; values above unity favour the horizon-mode hypothesis. The diamond and square markers at zero offset show the Bayes factors recovered from injections of the NRSur7dq4 and SEOBNRv5PHM waveforms, respectively.}
    \label{fig:bf}
\end{figure}

The Bayes factor $\mathcal{B}^{\rm HM}_{\rm OT}$ as a function of
the gate offset is shown in Fig.~\ref{fig:bf}. At the earliest gates the overtone hypothesis is strongly preferred, with $\mathcal{B}^{\rm HM}_{\rm OT}=(2.0^{+0.3}_{-0.2})\times10^{-3}$ and $(9\pm1)\times10^{-3}$ at $N=-6$ and $-4$. The Bayes factor crosses unity between $N=-4$ and $N=-2$, stays in the range $\mathcal{B}^{\rm HM}_{\rm OT}\simeq3$--$10$ for $-2\le N\le+2$, reaches a maximum of $\mathcal{B}^{\rm HM}_{\rm OT}=28.7_{-3.3}^{+3.7}$ at $N=+3$, and then drops to $4.4\pm0.5$ and $2.1\pm0.3$ at $N=+4$ and $+6$ before rising again to $8.8_{-1.0}^{+1.2}$ at $N=+8$, where little signal remains. Uncertainties on $\mathcal{B}$ are $1\sigma$ uncertainties obtained
by exponentiating $\ln\mathcal{B}\pm\sigma_{\ln\mathcal{B}}$, where
$\sigma_{\ln\mathcal{B}}$ is the quadrature sum of the
\texttt{dlog\_evidence} values reported by \textsc{Dynesty} for the two hypotheses; they therefore reflect the nested-sampling uncertainty only. 
The corresponding log-evidence curves for the two
hypotheses are shown in App.~\ref{app:evidence}
(Fig.~\ref{fig:evidence}).

The gate dependence of the Bayes factor is the key observational
result: the horizon-mode hypothesis is favoured over the overtone from $N=-2$ onward, most strongly at $N=+3$, while the overtone is strongly favoured at the earliest gates. This is the ordering expected from Sec.~\ref{sec:forced_obs}(b): while the source is still driving the exterior, the non-fundamental content of the signal is better described by a component near the horizon frequency than by a free $(2,2,1)$ mode. A zero-noise injection study using
\texttt{NRSur7dq4} and \texttt{SEOBNRv5PHM} waveforms, presented
in App.~\ref{app:injection} and Figs.~\ref{fig:final_state_horizonmode}, \ref{fig:final_state_overtone} and~\ref{fig:bf}, shows that under the overtone hypothesis the pipeline recovers the injected remnant spin and overestimates the injected mass by $3$--$4\,M_\odot$, that under the horizon-mode hypothesis both are shifted by amounts similar to those seen in the data, and that the preference for the horizon mode at
$N \sim 0$ is present in GR waveforms.

At the earliest gates ($N\le-4$) the ringdown segment includes the merger, and neither two-component model is adequate; the overtone template happens to absorb the pre-peak signal better, and we do not interpret these gates further. From $N=-2$ to $N=+3$ the horizon mode is preferred, with the Bayes factor largest at $N=+3$. Beyond that the preference drops ($\mathcal{B}^{\rm HM}_{\rm OT}\simeq4$ and $2$ at $N=+4$ and $+6$) and rises again at $N=+8$, where little signal remains. This is a pointwise scan over strongly overlapping data segments. The Bayes factors at different gates are correlated, the quoted uncertainties come from a single nested-sampling run per gate, and the maximum at $N=+3$ is not a calibrated global significance. We therefore do not extract a timescale from the scan. Taken at face value, the trend agrees with Sec.~\ref{sec:forced_obs}(b). It also agrees with the LVK analysis of GW250114, in which the fitted $(2,2,1)$ amplitude is consistent with free decay only for start times $\gtrsim6M_f$ after the peak, and adding a $(2,2,2)$ component extends the range of stable amplitudes down to $\approx3M_f$~\cite{LVK2026GW250114Spec}. In our reading, the residual isolated in GW250114 by QNM filtering~\cite{Lu2026GW250114Horizon,Chung2026DWAreaLaw} is the same driven content (Sec.~\ref{sec:forced_dw}). The injections of App.~\ref{app:injection} (markers in Figs.~\ref{fig:final_state_horizonmode}, \ref{fig:final_state_overtone}, and \ref{fig:bf}) show the same preference at $N=0$. The effect is therefore present in GR waveforms calibrated to NR, and it is not a sign of new physics.

\begin{figure*}[tp]
    \centering
    \includegraphics[width=0.9\textwidth]{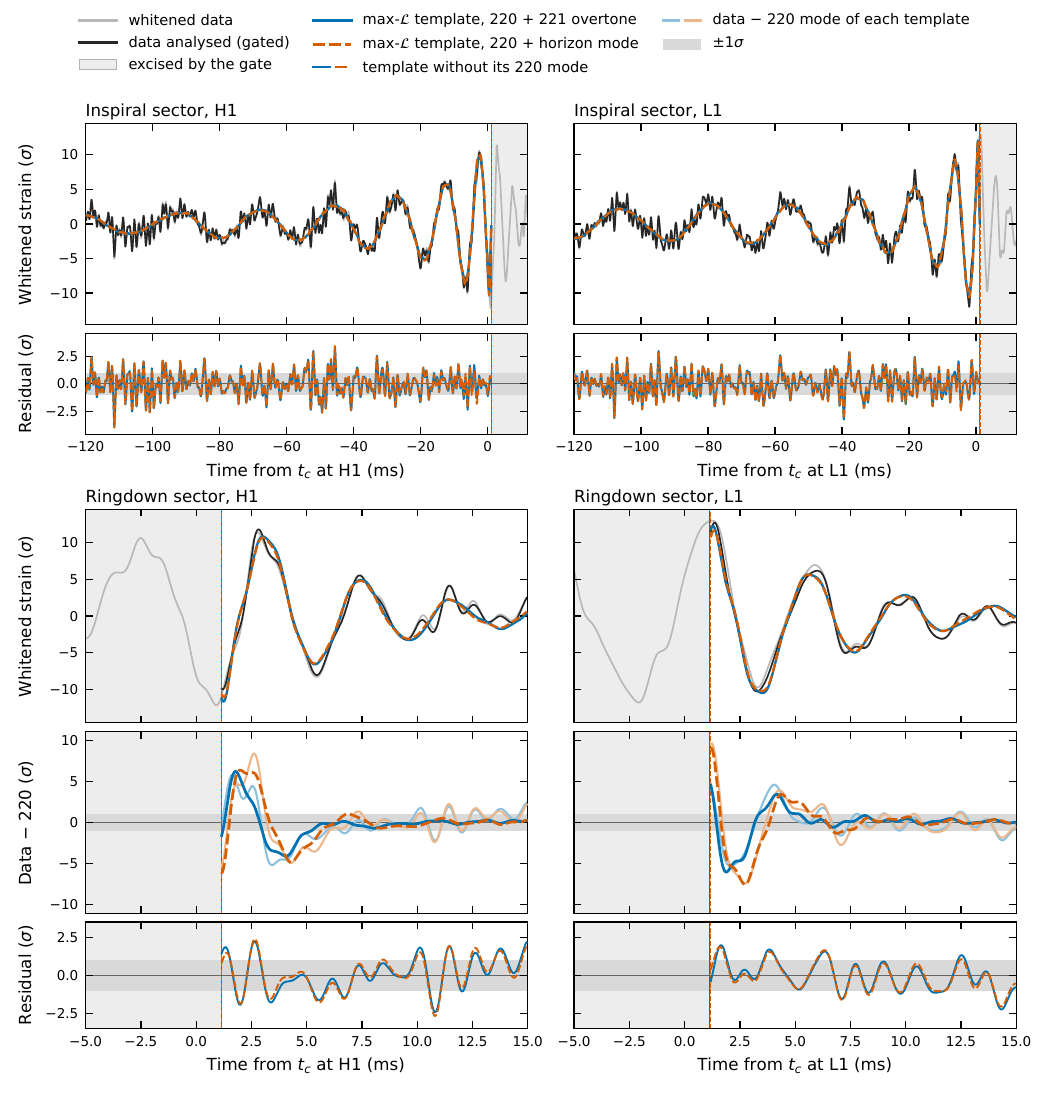}
    \caption{Whitened GW250114 data and the maximum-likelihood templates of the overtone hypothesis (220 + 221 overtone, solid blue) and of the horizon-mode hypothesis (220 + $(2,2,1)$ horizon mode, dashed orange) at gate offset $N=+3$, for the inspiral sector (top block) and the ringdown sector (bottom block) in H1 (left) and L1 (right). Time is measured from the coalescence time $t_c$ of the maximum-likelihood sample of the overtone analysis, at each detector. All series are whitened with the power spectral density (PSD) used in the analysis over $20$--$1024$\,Hz and normalised so that Gaussian noise has unit variance; the vertical axes are in units of the noise standard deviation $\sigma$. \textit{Whitened strain}: light grey, whitened data without gating; dark grey, whitened gated and in-painted data, the quantity that enters the gated likelihood; coloured, the templates, gated in the same way. The shaded interval is excised by the gate, which starts or ends at $t_c+NM_f$; vertical lines in the colour and line style of each template mark the gate edge of that hypothesis. \textit{Data $-$ 220} (ringdown sector only): light blue and light orange, the gated data with the $(2,2,0)$ mode of the overtone and of the horizon-mode template subtracted; blue and orange, the rest of each template, i.e.\ the overtone alone and the horizon mode alone. The $(2,2,0)$ mode subtracted is that of the two-component best fit of each hypothesis, not of a separate single-mode fit. \textit{Residual}: whitened gated data minus the full template. The grey bands mark $\pm1\sigma$. Because the likelihood marginalises over polarisation separately in each sector, each sector's template is drawn at the polarisation angle that maximises that sector's likelihood. The maximum-likelihood point is the highest-likelihood nested sample of each run.}
    \label{fig:maxl_waveforms}
\end{figure*}

Figure~\ref{fig:maxl_waveforms} compares the data with the maximum-likelihood templates of the two hypotheses at $N=+3$, where the Bayes factor peaks, and Fig.~\ref{fig:maxl_waveforms_N0} in App.~\ref{app:maxl} does the same at $N=0$. At both offsets the two templates follow the whitened data closely in the inspiral and in the ringdown sector, they are almost indistinguishable by eye, and the residuals in the plotted windows are consistent with Gaussian noise of unit variance (Table~\ref{tab:maxl_fit}). The second row of the ringdown block shows what the second component of each hypothesis fits: the data with the $(2,2,0)$ mode of that hypothesis' best fit subtracted, together with the rest of its template, i.e.\ the overtone or the horizon mode alone. In both hypotheses the second component is a short transient. It peaks at $6$--$9\sigma$ per sample at $N=+3$ and at $12$--$16\sigma$ at $N=0$, and it has fallen below $1\sigma$ within $4.0$--$6.1$\,ms ($N=+3$) and $3.5$--$8.5$\,ms ($N=0$) of the start of the analysed ringdown data, after which the data minus the $(2,2,0)$ mode are consistent with noise. The overtone and the horizon mode thus describe the same early transient, each relative to its own $(2,2,0)$ mode. The two components of a template partly cancel (Table~\ref{tab:maxl_ringdown}), so that each alone can be louder than their sum, and the split between them is not measured independently of the model. The figures therefore do not show a horizon mode resolved in the data. The preference between the hypotheses is quantified by the likelihood and the evidence: the horizon-mode hypothesis raises the maximum log-likelihood by $\Delta\ln\mathcal{L}_{\rm max}=2.34$ at $N=+3$ and $1.94$ at $N=0$, which is of the same size as $\ln\mathcal{B}^{\rm HM}_{\rm OT}=3.36$ and $1.83$ at these gates.

The displacement of the inferred remnant is part of the result, and we expect it. Under the horizon-mode hypothesis the ringdown sector returns a remnant that is heavier and spins faster than the inspiral predicts (Fig.~\ref{fig:final_state_horizonmode}), and it does so by a similar amount for both GR injections: at $N=0$ the paired shifts are $+5.7$ and $+7.8\,M_\odot$ in mass and $+0.085$ and $+0.107$ in spin, against $+8.2\,M_\odot$ and $+0.123$ in the data (App.~\ref{app:injection}). The overtone fit is displaced in the same direction in mass, by $1$--$3\,M_\odot$ less, but its spin shift is not significant before $N=+3$ and is absent in the injections (Fig.~\ref{fig:final_state_overtone}). The mass shift therefore belongs to the early signal, whichever template is fitted to it, while the spin shift is specific to the horizon mode. The reason is that the horizon mode is emitted while the remnant is still forming. Its frequency and damping are set by the angular velocity and surface gravity of the horizon at the time of emission, when the horizon is still growing and the exterior still holds energy and angular momentum that have yet to be radiated or absorbed. Nonlinear backreaction changes the mass and spin of the black hole during the ringdown~\cite{Sberna2022AIME,Zhu2024ChangingMassSpin}, so the mass and spin that enter the horizon mode are those at emission, not the final ones, whereas the inspiral sector predicts the final ones. The values of $(M_f,\chi_f)$ returned by the horizon-mode fit should therefore be read as effective parameters of the emitting horizon.

Two numbers make this concrete. Along the range of fitted values, from $(M_f,\chi_f)\approx(75\,M_\odot,0.78)$ to $(79\,M_\odot,0.84)$, the frequency of the $(2,2,0)$ mode stays within $2\%$ of its value at the remnant predicted by the inspiral, while the horizon mode oscillates $10$--$20\%$ faster and decays $17$--$29\%$ more slowly than it would at that remnant. A higher rotation rate and a lower damping rate are also what the plunging-particle model gives for a source that has not yet reached the horizon. In Eq.~\eqref{eq:omegaG} the decay rate lies below $\kappa$, and inside the light ring, where the source orbits faster than the horizon rotates, the real part lies above $m\Omega_{\rm H}$. The fitted frequency corresponds to $\beta\approx0.55$--$0.7$ and $\Omega\approx(0.23$--$0.26)/M_f$, between the horizon value $\Omega_{\rm H}=0.196/M_f$ and the light-ring value $0.277/M_f$ of a $\chi=0.68$ black hole. The shift is larger than a literal change of the Kerr parameters could be: the fitted mass exceeds the total mass of the binary, so it is not the mass of the space-time at the time of emission. It parametrizes, through the Kerr relations built into the template, the rotation rate and surface gravity of a horizon that is still dynamical.

That the injections reproduce the shift shows that it is a property of GR waveforms and not a departure from Kerr. It does not by itself show that the extra content is a horizon mode. A combination of ordinary Kerr components that the fitted model leaves out---the $(2,2,1)$ and $(2,2,2)$ modes together, higher multipoles, or quadratic modes---could shift the inferred parameters in a similar way, and the present two-template comparison cannot tell these apart.

Taken together, Figs.~\ref{fig:final_state_horizonmode}--\ref{fig:bf}
show that:
\begin{enumerate}
    \item under both hypotheses the ringdown sector returns a heavier remnant than the inspiral sector for $N\ge-1$, by $4$--$9\,M_\odot$ for the overtone and $5$--$11\,M_\odot$ for the horizon mode; the horizon mode also returns a higher spin, by $0.09$--$0.16$, at all $N\ge-2$, whereas under the overtone hypothesis the spin shift is significant only at $N=+3$, $+6$ and $+8$. The GR injections reproduce these shifts at $N=0$ (App.~\ref{app:injection} and Fig.~\ref{fig:final_state_horizonmode}), as expected for a signal emitted before the horizon has settled;
        \item the horizon-mode hypothesis is favoured over the overtone for all $N\ge-2$, with $\mathcal{B}^{\rm HM}_{\rm OT}$ between $2.1\pm0.3$ and $28.7^{+3.7}_{-3.3}$ (nested-sampling uncertainties only) and a maximum at $N=+3$, while the overtone is strongly favoured at the earliest gates.
\end{enumerate}

We leave several extensions for future work. The comparison should go beyond two models. On the same data and with the same priors, it should include the fundamental mode alone, the fundamental mode with both the first overtone and the horizon mode, models containing the $(2,2,2)$ mode, and the fundamental mode plus a damped sinusoid with free frequency and damping time, which would measure the second component directly. Each Bayes factor quoted here comes from a single nested-sampling run and should be checked with independent runs. The injection study should be repeated at every gate, with noise realizations, and with NR waveforms injected directly. The two remnant estimates should also be compared through the joint posterior of mass and spin, and not only through the one-dimensional paired differences used here. Finally, the same search can be carried out on other loud events of the current and next observing runs, where a larger signal-to-noise ratio would allow the frequency and decay rate of the second component to be measured directly.

\section{\label{Conclusion}Conclusion and discussion}
 Solutions of the black hole perturbation equations contain both ingoing and outgoing waves near the horizon, and the QNM boundary condition keeps only the former. We asked what local regularity requires of the outgoing waves, and what follows for the ringdown of a merger. Our conclusions are of three kinds, which we summarize separately.

\textit{Local regularity.} For a separated, scalar-like perturbation in a frame regular at the future horizon, bounded backreaction stress tensor, which depends on first derivatives, excludes outgoing waves with $\omega_I<\kappa$, but permits faster-damped ones (Table~\ref{tab:regularity}). Among the latter, smoothness selects the monomials in the Kruskal coordinate, the horizon modes, with frequencies $\omega=m\Omega_{\rm H}-ij\kappa$, $j\in\mathbb{N}$. At the rungs we examined, the ingoing solution is obstructed by a logarithm (Sec.~\ref{Horizon Modes}). For Schwarzschild black holes, and for the prograde $\ell=m=2$ branch of Kerr at all spins, only the fundamental QNM lies below the threshold (Figs.~\ref{fig:schwarzschild_fields} and \ref{fig:kerr_ladder}). These are, however, only local statements about near-horizon perturbations, which leave the QNM spectrum of the retarded Green's function, overtones included, unchanged. Their extension from a scalar-like variable to the gravitational field in a horizon-regular gauge is an assumption, which an explicit metric reconstruction should test.

\textit{The forced response.} A source falling through the horizon approaches it at the rate $\kappa$, up to a logarithmic prefactor that we have computed for geodesic plunges (Table~\ref{tab:plunge}), and close to the horizon it acts on the exterior at the horizon frequencies. Earlier in the plunge its forcing decays more slowly than $\kappa$ [Eq.~\eqref{eq:omegaG}]; once the source is behind the potential barrier, the forcing of the gravitational waveform decays at $3\kappa$~\cite{Nakamoto2026PoleSplitting,Laeuger2026SourceImprints}. The response of each QNM pole has a driven and a free part, and the free part dominates only once the forcing decays faster than the mode (Sec.~\ref{sec:forced_which}). The fundamental mode therefore separates from the source early, in any description of the source, whereas the overtones are driven until the source is screened and ring freely only afterwards, with whatever amplitude they have left. Once the source is close enough to the horizon that it acts only at the horizon frequencies, the driven parts of the individual modes cancel one another, and the waveform is a sum of QNMs with constant amplitudes~\cite{Weller2026DrivenToFree,Kuntz2026RedshiftModes}. The cancellation results of Refs.~\cite{Kuntz2026RedshiftModes,Kubota2026HorizonCancel} rule out a stationary horizon line at infinity. They do not rule out a transient near the horizon frequency, which is how the direct wave is now understood (Sec.~\ref{sec:forced_dw}). Whether nonlinear backreaction lets the $j=1$ rung reach infinity is an open question, which the linear cancellation results do not settle.

\textit{Identifiability in data.} Overtone amplitudes fitted before the source is screened cannot in general be read as amplitudes of free oscillations. Whether an overtone can be identified afterwards depends on the amplitude it has left, the signal-to-noise ratio, the covariance with other components and the template error, and has to be decided case by case (Sec.~\ref{sec:forced_which}). In GW250114 (Sec.~\ref{sec:horizon}) a model with the fundamental mode and the horizon mode $2\Omega_{\rm H}-i\kappa$ is preferred over the fundamental mode plus its first overtone for start times from $2M_f$ before to $8M_f$ after coalescence, most strongly at $3M_f$. The horizon-mode fit returns a remnant that is heavier and spins faster than the inspiral predicts, as expected for a mode emitted before the horizon has settled, and zero-noise injections of NR-calibrated waveforms reproduce both the preference and the shift. The overtone fit also returns a heavier remnant, by a smaller amount, but no comparable spin shift. We read this as evidence that the early non-fundamental content of the ringdown is better described by a driven component near the horizon frequency than by a free first overtone. The evidence is not yet conclusive: only two models have been compared, each Bayes factor comes from a single nested-sampling run, and the injections cover one start time without noise. The extensions that would address this are listed at the end of Sec.~\ref{sec:horizon}. The three hypotheses of Sec.~\ref{sec:forced_obs} are the tests that would establish the mechanism.

Compared with the forced-excitation studies of Refs.~\cite{Weller2026DrivenToFree,DeAmicis2026DynQNM,DeAmicis2026DynQNMII}, what is new here is the link between the local regularity threshold and the rate at which a horizon-crossing source settles, the identification of the horizon monomials with the frequencies at which it acts, and the observation that the fundamental modes are the only ones that lie below every rate the horizon imposes on the source. In our reading, the signatures of horizon dynamics reported in the direct wave~\cite{Oshita2025DirectWaves,Lu2026GW250114Horizon} are the driven, non-fundamental content of this transient. Finding the horizon monomials directly would require a horizon-penetrating extraction of the near-horizon field in numerical simulations, where the field of the settling remnant should be a series in the Kruskal coordinate.

\textit{Outlook.} Three lines of work would show how far this picture holds. On the theory side, the regularity threshold should be derived for metric perturbations in a horizon-regular gauge, and the forcing should be computed at second order, where the stress of the ingoing waves is itself a source that crosses the horizon~\cite{Sberna2022AIME}. That calculation would decide whether the $j=1$ rung, which cancels in linear theory, reaches infinity in a comparable-mass merger. In numerical relativity, the hypotheses of Sec.~\ref{sec:forced_obs} can be tested on existing waveform catalogs, by measuring how the onset of free ringing scales with $1/\kappa$ across remnant spins, and the mass and spin returned by a horizon-mode fit can be compared with the angular velocity and surface gravity of the dynamical horizon in the same simulation. On the observational side, GW250114 is a single event at a single spin. Louder signals in the coming observing runs, and much louder ones in next-generation detectors, should allow the frequency and decay rate of the early non-fundamental component to be measured directly, without tying them to a template. If the picture holds, it has two consequences. Tests of general relativity that rely on overtones should start after the source is screened, or model the driven component explicitly. The first few $M$ after the peak would then record how the remnant horizon forms and settles, which the late-time QNMs, fixed by the final mass and spin alone, cannot do.

\begin{acknowledgments}

We thank Bruce Allen, Emanuele Berti, Vitor Cardoso, Badri Krishnan, Sizheng Ma, Alex Nielsen, José Luis Jaramillo, Rodrigo Panosso Macedo, and Aaron Zimmerman
for helpful comments and discussions. J.A. thanks Collin Capano and Alex Correia for their development of the hierarchical model and gating in PyCBC, and for their suggestions during a parallel project that were useful for the GW250114 data analysis in this work. 

We finally thank the late Rafael Sorkin for bringing the original idea to our attention and subsequent discussions and comments. This work is dedicated to his memory.

This research was supported in part by Perimeter Institute for Theoretical Physics. Research at Perimeter Institute is supported in part by the Government of Canada through the Department of Innovation, Science and Economic Development and by the Province of Ontario through the Ministry of Colleges and Universities. N.A. is supported by the Natural Sciences and Engineering Research Council of Canada (NSERC). 

Quasinormal-mode frequencies were computed with the \texttt{qnm} package~\cite{Stein2019qnm}. The GW250114 analysis used \texttt{PyCBC Inference}~\cite{Biwer2019} and \texttt{dynesty}~\cite{Speagle2020}. This research has made use of data or software obtained from the Gravitational Wave Open Science Center (gwosc.org)~\cite{LVK2026OpenDataO4b}, a service of the LIGO Scientific Collaboration, the Virgo Collaboration, and KAGRA. This material is based upon work supported by NSF's LIGO Laboratory which is a major facility fully funded by the National Science Foundation, as well as the Science and Technology Facilities Council (STFC) of the United Kingdom, the Max-Planck-Society (MPS), and the State of Niedersachsen/Germany for support of the construction of Advanced LIGO and construction and operation of the GEO600 detector. Additional support for Advanced LIGO was provided by the Australian Research Council. Virgo is funded, through the European Gravitational Observatory (EGO), by the French Centre National de Recherche Scientifique (CNRS), the Italian Istituto Nazionale di Fisica Nucleare (INFN) and the Dutch Nikhef, with contributions by institutions from Belgium, Germany, Greece, Hungary, Ireland, Japan, Monaco, Poland, Portugal, Spain. KAGRA is supported by Ministry of Education, Culture, Sports, Science and Technology (MEXT), Japan Society for the Promotion of Science (JSPS) in Japan; National Research Foundation (NRF) and Ministry of Science and ICT (MSIT) in Korea; Academia Sinica (AS) and National Science and Technology Council (NSTC) in Taiwan.

N.A. acknowledges the use of Claude Fable (Anthropic) for checking calculations, surveying the literature, and drafting and revising the text, and of ChatGPT Astra (OpenAI) for independent reviews of the manuscript. The authors are responsible for the content of the paper.
\end{acknowledgments}

\section*{Data availability}
The data that support the findings of this article are openly available. The strain data of GW250114 are distributed by the Gravitational Wave Open Science Center~\cite{LVK2026OpenDataO4b}, and the posterior samples used to construct the \texttt{NRSur7dq4} injection are part of the GWTC-5.0 data release~\cite{GWTC5PE}. The configuration files, posterior samples and scripts of the analysis of Sec.~\ref{sec:horizon} are available in Ref.~\cite{jabedi_gw250114_pipeline}.

\appendix

\section{Injection study at the coalescence time}
\label{app:injection}

To validate the pipeline used in Sec.~\ref{sec:horizon} and App.~\ref{app:method}
and to confirm that the Bayesian model selection between the
horizon-mode and overtone hypotheses is not driven by a bias in the
recovery of the remnant parameters, we performed a set of zero-noise
injection studies at the gate offset $N = 0$, i.e.\ with the
ringdown gate anchored exactly at the coalescence time
$t_c$ (Eq.~\eqref{eq:gate}). Two independent injections were used,
differing in the waveform model adopted to generate the injected
signal:
\begin{itemize}
    \item an injection generated with \texttt{NRSur7dq4}, a
    numerical-relativity surrogate calibrated to a large catalog of
    NR simulations~\citep{Varma2019}, with component masses
    $(m_1, m_2) = (37.50, 34.50)\,M_\odot$, spin magnitudes of
    $0.03$ and $0.002$, and a luminosity distance of
    $409\,\mathrm{Mpc}$;
    \item an injection generated with \texttt{SEOBNRv5PHM}, the
    state-of-the-art effective-one-body model with higher modes and
    precession~\citep{Pompili2023,RamosBuades2023}, with
    $(m_1, m_2) = (34.79, 34.50)\,M_\odot$ and
    spin magnitudes of $0.51$ and $0.74$. The two spins lie close to
    the orbital plane, where they point in opposite directions; the
    effective aligned spin is $\chi_{\rm eff}=-0.12$ and the effective
    precession spin is $\chi_p=0.70$.
\end{itemize}
Both injections correspond to the same binary black hole
coalescence GW250114 analysed in this work, and their intrinsic
parameters are those quoted above. The NRSur7dq4 injection uses the maximum-likelihood sample of the NRSur7dq4
posterior for GW250114 from the GWTC-5.0 parameter estimation data
release~\citep{GWTC5PE}; the SEOBNRv5PHM injection uses the maximum-likelihood
sample of an independent SEOBNRv5PHM IMR analysis performed for this work.
The two injections therefore have different initial parameters. The injected
remnant mass and spin shown in Figs.~\ref{fig:final_state_horizonmode} and~\ref{fig:final_state_overtone} are obtained by applying the final-state mapping of
App.~\ref{app:mapping} to the injected parameters. The full parameter sets are tabulated in the analysis
repository~\citep{jabedi_gw250114_pipeline}. 

For each injection we ran two independent analyses, one assuming
the horizon-mode hypothesis (220 + $221_{\rm H}$, with $(\ell,m,n)=(2,2,1)$) and one assuming the
overtone hypothesis (220 + 221), using the same configuration,
priors, and sampler settings as in App.~\ref{app:method}. We then
computed the log Bayes factor
$\log\mathcal{B}^{\rm HM}_{\rm OT}$ between the two hypotheses,
defined in Sec.~\ref{sec:horizon}, and compared the ringdown inference of the
remnant mass and spin with the prediction obtained by mapping the
inspiral posterior through the final-state fits, as
described in App.~\ref{app:mapping}.

Figures~\ref{fig:final_state_horizonmode}, \ref{fig:final_state_overtone} and~\ref{fig:bf} summarise the results. They show the final mass and the final spin inferred from the inspiral sector and from the ringdown sector, for each of the two hypotheses (220 + Horizon
mode, and 220 + 221) and for each injected waveform. The
dashed and dotted horizontal lines mark the injected values,
obtained by applying the same final-state mapping to the injected parameters. Three features are worth highlighting:
\begin{enumerate}
        \item Under the overtone hypothesis the ringdown sector recovers the injected spin for both waveforms (paired shift between the ringdown and inspiral sectors of $+0.008$ in both cases, consistent with zero). The injected mass is recovered within the 90\% interval for \texttt{NRSur7dq4} ($71.5\,M_\odot$ against an injected $68.6\,M_\odot$; paired shift $+3.1\,M_\odot$, 90\% interval $-0.7$ to $+5.5$) but not for \texttt{SEOBNRv5PHM} ($70.1\,M_\odot$ against $66.2\,M_\odot$; paired shift $+3.8\,M_\odot$, 90\% interval $+0.4$ to $+7.7$).
    \item Under the horizon-mode hypothesis the ringdown sector recovers $M_f=74.0\,M_\odot$ and $\chi_f=0.768$ for \texttt{NRSur7dq4} (injected $68.57\,M_\odot$ and $0.681$), and $M_f=74.3\,M_\odot$ and $\chi_f=0.758$ for \texttt{SEOBNRv5PHM} (injected $66.16\,M_\odot$ and $0.645$), in both cases with the injected values outside the 90\% credible intervals, while the inspiral sector recovers them (Figs.~\ref{fig:final_state_horizonmode},~\ref{fig:final_state_overtone}). The paired shifts between the ringdown and inspiral sectors are $+5.7\,M_\odot$ (90\% interval $+3.0$ to $+8.5$) and $+0.085$ ($+0.042$ to $+0.124$) for \texttt{NRSur7dq4}, and $+7.8\,M_\odot$ ($+3.9$ to $+10.3$) and $+0.107$ ($+0.046$ to $+0.146$) for \texttt{SEOBNRv5PHM}, against $+8.2\,M_\odot$ and $+0.123$ in the data at the same gate. The choice of secondary mode therefore does change the reconstruction of the remnant parameters, by an amount similar to the shift seen in the data; this is the shift discussed in Sec.~\ref{sec:horizon}.
    \item The uncertainties on $M_f$ and $\chi_f$ are comparable
    between the NRSur7dq4 and SEOBNRv5PHM injections. Because the two
    injections also differ in their intrinsic parameters, this does
    not by itself establish that the final-state measurement is
    independent of the waveform family
    (the analysis itself always uses \texttt{SEOBNRv5PHM} for the
    inspiral submodel).
\end{enumerate}

The markers at zero offset in Fig.~\ref{fig:bf} show the Bayes factor
$\mathcal{B}^{\rm HM}_{\rm OT}$ for the two injections. Both
injections yield $\log\mathcal{B}^{\rm HM}_{\rm OT} > 0$, i.e.\ a
preference for the horizon-mode hypothesis over the overtone
at $N = 0$, with
$\log\mathcal{B}^{\rm HM}_{\rm OT} = +1.6$ for NRSur7dq4 and
$+4.7$ for SEOBNRv5PHM. These values correspond to Bayes factors
of $\mathcal{B}^{\rm HM}_{\rm OT} \simeq 5$ and $\simeq 110$,
respectively: a modest preference for the first injection and a strong one for the second. Both injections prefer the horizon-mode template at $N=0$, as the data do. The factor of about $20$ between the two Bayes factors shows a substantial dependence on the injected signal. Since the two injections differ in their intrinsic parameters as well as in the waveform model (the \texttt{SEOBNRv5PHM} sample has large in-plane spins and the \texttt{NRSur7dq4} sample almost none), the comparison does not isolate the effect of the waveform family. Neither waveform model has a horizon mode put in by hand: \texttt{NRSur7dq4} interpolates NR waveforms and \texttt{SEOBNRv5PHM} is calibrated to them. The preference therefore reflects content that GR waveforms themselves contain near the peak, which at this gate the horizon-mode template describes better than the overtone template does. 

The injection study therefore shows that the preference for the horizon-mode template at $N\sim0$, and the accompanying shift of the inferred remnant under that template, are both present in GR waveforms calibrated to NR. It is a zero-noise study at a single gate: it exposes a possible systematic bias of the template but does not establish calibrated statistical coverage, nor the gate dependence of the preference, which require injections at every gate with noise realizations.

\section{Analysis pipeline and priors}
\label{app:method}

\subsection{Data and Bayesian inference setup}

We analyse GW250114 with the \texttt{PyCBC Inference}
framework~\citep{Biwer2019}, using 4\,kHz strain data from H1 and L1~\cite{LVK2026OpenDataO4b}
around the GPS trigger time $1420878141.223\,\mathrm{s}$. The data are
downsampled to 2048\,Hz, high-pass filtered at 15\,Hz, and the PSD is
estimated with the median-mean method over 512\,s of data
($-256\,$s to $+256\,$s around the trigger) in 8\,s segments with a
4\,s stride. We adopt a gated Gaussian likelihood for both detectors
and both submodels, which analytically marginalises over the
polarization angle; the likelihood uses a low-frequency cutoff of
20\,Hz, while the waveform is generated from 13\,Hz. The analysis uses
the LVK-calibrated strain products directly, with no marginalisation
over calibration uncertainty. The gate windows, the likelihood
implementation, and the pinned PyCBC version are documented in the
analysis repository~\cite{jabedi_gw250114_pipeline}. 

The posterior is sampled with the nested sampler
\texttt{dynesty}~\citep{Speagle2020} using \texttt{nlive}=8000
live points and a stopping criterion
\(\Delta\log\mathcal{Z}=0.1\). The stopping criterion controls the
termination of the run and is not an evidence uncertainty: the
\(1\sigma\) numerical uncertainty on each log evidence is the
\texttt{dlog\_evidence} value reported by \textsc{Dynesty}, and the
uncertainty on the log Bayes factor is
\(\sigma_{\ln\mathcal{B}}=
\sqrt{\sigma_{\rm HM}^{2}+\sigma_{\rm OT}^{2}}\). For the runs used
here \(\sigma_{\ln Z}\approx0.08\)--\(0.09\) per hypothesis, giving
\(\sigma_{\ln\mathcal{B}}\approx0.12\)--\(0.13\) at every gate; these
are the uncertainties quoted in Sec.~\ref{sec:horizon}. Only one
nested-sampling run was performed per gate, so no run-to-run scatter is
quoted, and the uncertainties reflect the internal \textsc{Dynesty}
error only. 

\subsection{Hierarchical inspiral--ringdown model}

We decompose the signal into two submodels connected
by a set of shared parameters:
\begin{itemize}
    \item an \emph{inspiral} submodel with the
    \texttt{SEOBNRv5PHM} approximant~\citep{Pompili2023,RamosBuades2023}, gated on
    from $t_{\rm gate}^{\rm ins}$ to the start of the analysis
    segment, and tapered with a constant window of
    $0.1\,\mathrm{s}$ at the start of the gate;
    \item a \emph{ringdown} submodel described by time-domain QNMs
    with \texttt{TdQNMfromFinalMassSpin}, gated on from
    $t_{\rm gate}^{\rm rd}$ to $t_{\rm gate}^{\rm ins}$.
\end{itemize}
The two submodels share the sky position $(\alpha,\delta)$, the
coalescence time $t_c$, the luminosity distance, and the
inclination through a hierarchical waveform transform, while the
intrinsic parameters (masses, spins, QNM amplitudes and phases)
are sampled independently. Because the extrinsic parameters are shared and the noise in the two segments is correlated, the remnant estimates from the two sectors are not statistically independent.



For the inspiral submodel, the component masses are drawn from a joint
uniform prior in $m_1$ and $m_2$, corresponding to
$\mathcal{M}_c \in [29,33]\,M_\odot$ and $q \in [1,3]$; the two
dimensionless spin magnitudes are uniform over $a_{1,2}\in[0,0.99]$
with isotropic orientations on the sphere; the luminosity distance is
uniform in comoving volume over $200$--$580\,\mathrm{Mpc}$; the
inclination follows a $\sin\iota$ prior; and the coalescence time is
uniform within $\pm 50\,\mathrm{ms}$ of the trigger.

For the ringdown submodel, the remnant mass and spin have uniform
priors $M_f \in [10,200]\,M_\odot$ and
$\chi_f \in [-0.99,0.99]$. The amplitude of the dominant $(2,2,0)$
mode is log-uniform over $[10^{-3},5]$, while the amplitude of the second component is defined relative to $A_{220}$ and drawn uniformly over $[0,5]$; the phases of both components are uniform, with the subdominant phase measured relative to $\phi_{220}$. The amplitudes are defined at the start of the ringdown segment, $t_{\rm gate}^{\rm rd}$ of Eq.~\eqref{eq:gate}. 
The complete
prior specification, the full inference configuration, and the
posterior samples are distributed with the analysis code
\cite{jabedi_gw250114_pipeline}.

\subsection{Inspiral-to-remnant mapping}
\label{app:mapping}

To compare the ringdown measurement of $(M_f, \chi_f)$ with the
inspiral prediction we map each posterior sample of the inspiral
parameters $(\mathcal{M}_c, q, \mathbf{s}_1, \mathbf{s}_2)$ to a
remnant using the final-state
fits implemented as
\texttt{final\_mass\_from\_initial} and
\texttt{final\_spin\_from\_initial} in \texttt{PyCBC}. 
Component masses are recovered from $(\mathcal{M}_c, q)$, and the spherical
spin components $(a_i, \theta_i, \varphi_i)$ are converted to
Cartesian vectors before evaluating the fits. This propagates the
full inspiral posterior to the remnant, so the comparison with
the ringdown posterior is performed at the level of the full
distributions rather than point estimates.

\subsection{Evidence curves}
\label{app:evidence}

\begin{figure}[t!]
    \centering
    \includegraphics[width=0.48\textwidth]{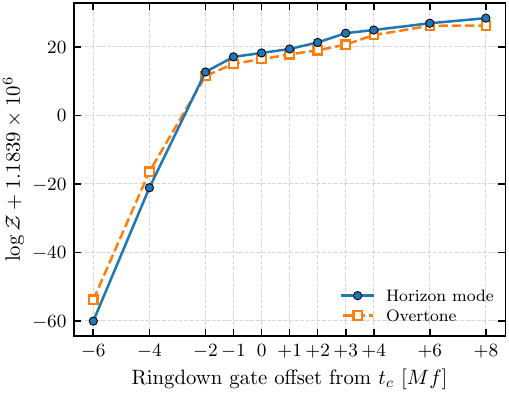}
    \caption{Log Bayesian evidence of the horizon-mode (blue
    circles) and overtone (orange squares) hypotheses as a
    function of the ringdown gate offset from $t_c$, in units of
    the final mass.}
    \label{fig:evidence}
\end{figure}

Figure~\ref{fig:evidence} shows the log evidence of both
hypotheses as a function of the gate offset. For both models the
log evidence increases as the gate is pushed to later times. Moving the gate changes which samples are analysed, so the absolute values of the evidence at different gates refer to different data, and their increase is not by itself a sign of improved modelling or of increased signal-to-noise ratio. Only their difference at a given gate, the Bayes factor of Fig.~\ref{fig:bf}, is meaningful.

\subsection{Maximum-likelihood waveforms and residuals}
\label{app:maxl}

\begin{figure*}[tp]
    \centering
    \includegraphics[width=0.9\textwidth]{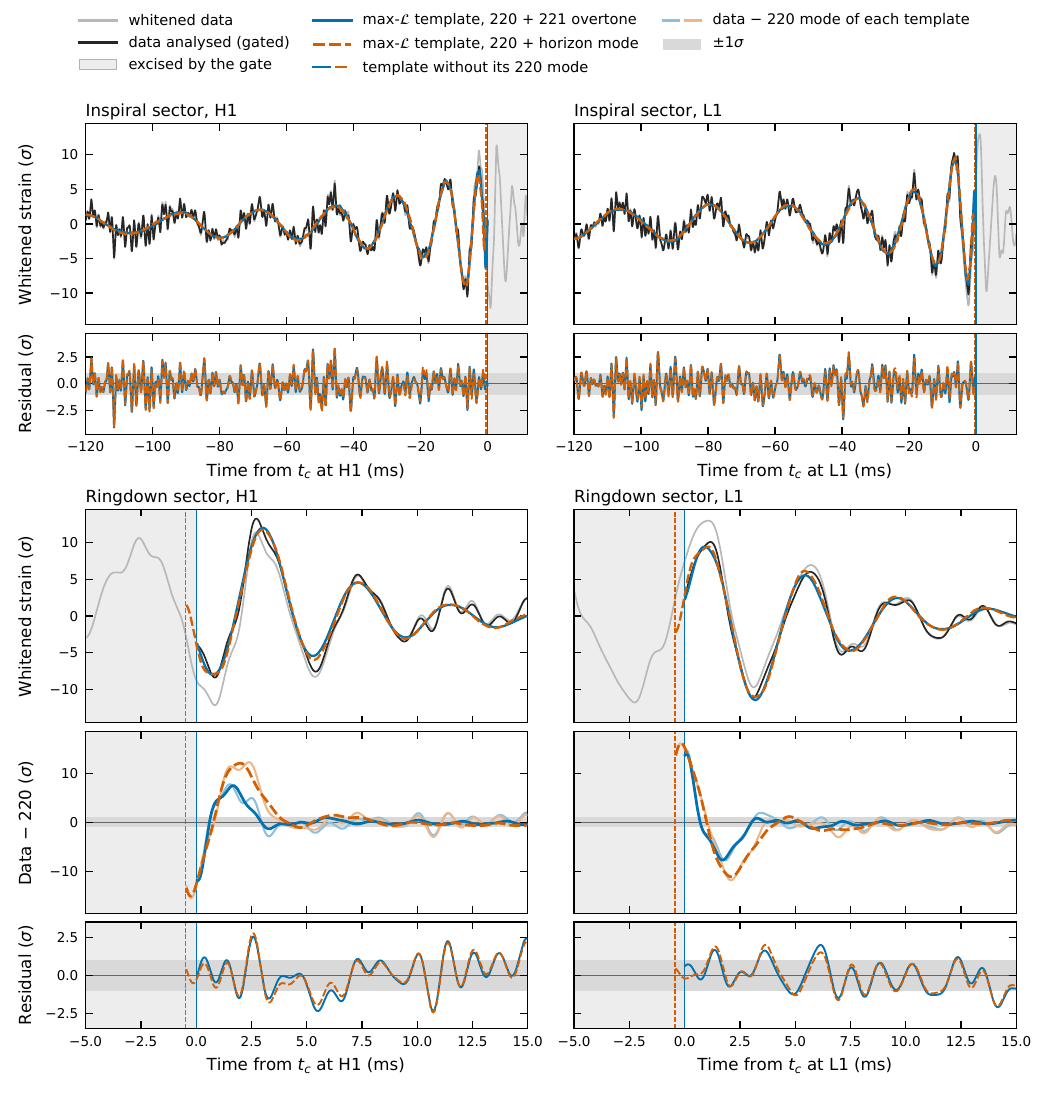}
    \caption{Same as Fig.~\ref{fig:maxl_waveforms} for gate offset $N=0$, where the gate edge is the coalescence time $t_c$ of each hypothesis. The time origin and the shaded interval are those of the overtone analysis. The gate edge of the horizon-mode maximum-likelihood point (dashed orange vertical line) is $0.48$\,ms earlier in H1 and $0.43$\,ms earlier in L1.}
    \label{fig:maxl_waveforms_N0}
\end{figure*}

\textit{Reconstruction.} The waveforms in Figs.~\ref{fig:maxl_waveforms} and~\ref{fig:maxl_waveforms_N0} are reconstructed from the posterior files of the $N=+3$ and $N=0$ analyses. For each analysis we take the nested sample with the largest likelihood, rebuild the hierarchical model from the configuration stored with the run~\cite{jabedi_gw250114_pipeline}, and evaluate it at that sample. The recomputed log-likelihoods agree with the stored values to better than $2\times10^{-6}$ in three of the four runs and to $2\times10^{-3}$ in the fourth ($N=0$, overtone), so the reconstruction uses the same data, PSDs, gates and waveform models as the sampling. For each sector and detector we form the template at the polarisation angle that maximises that sector's likelihood, and gate and in-paint the data and the template with the same operation used in the likelihood~\cite{Zackay2021,Capano2021}. Data, templates and residuals are whitened by multiplying their Fourier transforms by $S_n^{-1/2}(f)$ over $20$--$1024$\,Hz and normalised so that stationary Gaussian noise has unit variance per sample. With this normalisation the sum over the segment of the squared whitened gated residual equals $-2\ln\mathcal{L}$ of that sector and detector, up to the parameter-independent normalisation; it reproduces the maximum-polarisation log-likelihood of each sector.

\textit{The second component.} For the ringdown sector we also generate the template with all mode amplitudes except $A_{220}$ set to zero, at the same polarisation angle, and subtract it from the gated data (second row of the ringdown block in the figures). The remaining part of the template is the second component alone. The maximum-likelihood ringdown parameters and the signal-to-noise ratios (SNRs) of the two components are listed in Table~\ref{tab:maxl_ringdown}; the remnant parameters lie inside the ranges of the ringdown-sector posteriors quoted in Sec.~\ref{sec:horizon}. In every case the two components are anticorrelated, with overlaps between $-0.10$ and $-0.64$, so the SNR of the full ringdown template is smaller than the quadrature sum of its parts. The cancellation is strongest for the horizon-mode template at $N=0$, where the $(2,2,0)$ mode and the horizon mode alone have network SNRs of $46.3$ and $40.8$ and their sum has $37.2$. The $(2,2,0)$ mode that is subtracted belongs to the two-component fit of each hypothesis. The residual power of the data minus that mode, listed in the last rows of Table~\ref{tab:maxl_fit}, therefore measures how much of the template is carried by the second component in that fit, and is larger than what a re-fitted single-mode model would leave. It cannot be compared between the hypotheses as a measure of evidence.

\textit{Comparing sectors between hypotheses.} For each sector and detector, and at fixed polarisation, the log-likelihood is the sum of a term that depends only on the data which the sector analyses and a term that depends on the template,
\begin{equation}
    \ln\mathcal{L}=-\tfrac12\langle d|d\rangle+\left[\langle d|h\rangle-\tfrac12\langle h|h\rangle\right],
    \label{eq:lnL_split}
\end{equation}
where $\langle\cdot|\cdot\rangle$ is the gated inner product. Table~\ref{tab:maxl_split} lists the differences of the two terms between the maximum-likelihood points of the two hypotheses. The entries for a single sector are larger than their sum over the sectors, at $N=0$ by more than an order of magnitude: there the data term of the inspiral sector differs by $+48.4$ and its template term by $-49.3$, while the log-likelihood summed over both sectors differs by $1.5$. The reason is that the boundary between the sectors, $t_c+NM_f$ at each detector, is a sampled quantity and differs between the two points. At $N=+3$ the boundary of the horizon-mode point is later by $8\,\mu$s in H1 and $33\,\mu$s in L1 ($0.016$ and $0.068$ samples), of which $31\,\mu$s comes from its larger $M_f$ and the rest from the arrival times. At $N=0$, where the boundary is $t_c$ itself, it is earlier by $0.48$\,ms in H1 and $0.43$\,ms in L1 ($0.99$ and $0.89$ samples). The signal is loud at the boundary (the whitened data reach $11$--$13\sigma$ there at $N=+3$), so even these displacements move a significant amount of signal from one sector to the other. The gate is not tied to the sampling grid: the data are first shifted by the sub-sample part of the gate time, which is a band-limited interpolation, whole samples are then excised and in-painted, and the shift is undone. The excised interval and the likelihood therefore vary continuously with the gate time. This is seen directly at $N=+3$, where the two boundaries lie within the same sample in each detector and the data terms nevertheless differ, by $-0.8$ (H1) and $-5.2$ (L1) in the inspiral sector and by $+1.1$ and $+3.5$ in the ringdown sector. Quantities that refer to a single sector, such as the SNR of the ringdown template, therefore cannot be compared between the hypotheses. At $N=+3$ the ringdown template of the horizon-mode hypothesis is quieter than that of the overtone hypothesis ($\rho=34.5$ against $35.6$, Table~\ref{tab:maxl_ringdown}) and its template term is lower by $2.5$, while the hypothesis is preferred: its ringdown sector starts later and analyses less of the loud signal at the boundary, which its inspiral sector analyses instead. Only the likelihood summed over both sectors, and the evidence, compare the hypotheses on the same data.

\textit{Segment ends.} The value of $\ln\mathcal{L}$ itself contains a large contribution that is common to both hypotheses. The likelihood is evaluated on a $24$\,s segment, the $16$\,s analysis window padded by $4$\,s on each side, which is Fourier-transformed without a window. At the time of the event the H1 strain in this segment is dominated by narrow spectral lines, the strongest at $508$\,Hz and $989$\,Hz with amplitudes of $9.4\times10^{-21}$ and $7.1\times10^{-21}$, about $30$ times the r.m.s.\ of the broadband noise between $20$ and $490$\,Hz. These lines are not periodic over the segment, and after whitening the mismatch between the end and the start of the segment appears as excess power confined to the segment ends. Of the summed squared whitened residual in H1, $1.13\times10^{6}$ for $4.7\times10^{4}$ analysed samples, $95\%$ lies in the first and the last second of the segment, which are more than $17$\,s before and $4.8$\,s after the merger. Between $2$\,s and $22$\,s from the start of the segment the residual power per sample lies between $0.90$ and $1.10$ in every $1$\,s block, against an expected $0.98$. In L1 the strongest line in the analysed band is $16$ times weaker, and the first and last second contribute $22\%$ of $5.4\times10^{4}$. Tapering only the outermost $0.5$\,s at each end of the segment, as a test, lowers the sum by $9.9\times10^{5}$ (H1) and $7.2\times10^{3}$ (L1) for every sector, hypothesis and offset, the same to within $4$ and $0.4$. The part that depends on the hypothesis is small: at the maximum-likelihood points the difference $\ln\mathcal{L}_{\rm HM}-\ln\mathcal{L}_{\rm OT}$ is larger as analysed than with tapered ends by $0.12$ at $N=+3$ and by $0.26$ at $N=0$. It arises mainly in L1, and the same test on simulated Gaussian noise without lines gives $0.11$, so it is not caused by the H1 lines. This is $5\%$ and $13\%$ of $\Delta\ln\mathcal{L}_{\rm max}$, and one to two times the nested-sampling uncertainty of $\ln\mathcal{B}^{\rm HM}_{\rm OT}$; its effect on the evidence was not evaluated.

\textit{Residuals in the plotted windows.} Table~\ref{tab:maxl_fit} lists the network matched-filter SNR of each sector and the sum of squared whitened residuals over the analysed samples inside the plotted windows. In every window the sum is consistent with the number of samples, with tail probabilities between $0.20$ and $0.78$ under a $\chi^2$ distribution. This count does not subtract the fitted parameters and treats neighbouring whitened samples as independent, which holds only approximately next to the gate, so it is a consistency check rather than a goodness-of-fit test.

\begin{table}[t]
\caption{Ringdown sector at the maximum-likelihood points. $f$ and $\tau$ are the frequency and damping time of the $(2,2,0)$ mode and of the second component (the $(\ell,m,n)=(2,2,1)$ overtone, OT, or the $(\ell,m,j)=(2,2,1)$ horizon mode, HM), evaluated at the listed detector-frame $M_f$ and $\chi_f$; $A_2/A_{220}$ is the sampled amplitude ratio. $\rho_{220}$ and $\rho_2$ are the network SNRs (template norms with the gate applied, H1 and L1 in quadrature) of the $(2,2,0)$ mode alone and of the second component alone, $\rho$ that of the full ringdown template, and $\mathcal{O}$ the normalised overlap between the two components in H1 and L1.}
\label{tab:maxl_ringdown}
\begin{ruledtabular}
\begin{tabular}{lcccc}
 & \multicolumn{2}{c}{$N=+3$} & \multicolumn{2}{c}{$N=0$}\\
 & OT & HM & OT & HM\\
\hline
$M_f$ [$M_\odot$] & 76.6 & 78.7 & 73.9 & 76.4\\
$\chi_f$ & 0.784 & 0.836 & 0.733 & 0.800\\
$f_{220}$ [Hz] & 242.9 & 250.9 & 239.6 & 247.9\\
$\tau_{220}$ [ms] & 4.92 & 5.33 & 4.58 & 4.98\\
$f_2$ [Hz] & 239.2 & 221.5 & 235.0 & 211.5\\
$\tau_2$ [ms] & 1.63 & 2.19 & 1.52 & 2.01\\
$A_2/A_{220}$ & 1.28 & 1.62 & 0.81 & 1.45\\
\hline
$\rho_{220}$ & 39.8 & 30.1 & 43.2 & 46.3\\
$\rho_2$ & 15.0 & 20.4 & 24.8 & 40.8\\
$\mathcal{O}$ (H1) & $-0.52$ & $-0.12$ & $-0.49$ & $-0.64$\\
$\mathcal{O}$ (L1) & $-0.39$ & $-0.10$ & $-0.56$ & $-0.64$\\
$\rho$ & 35.6 & 34.5 & 36.7 & 37.2\\
\end{tabular}
\end{ruledtabular}
\end{table}

\begin{table*}[t]
\caption{Fit quality of the maximum-likelihood templates. $\rho_{\rm mf}=\langle d|h\rangle/\langle h|h\rangle^{1/2}$ is the network matched-filter SNR of each sector with the gate applied (H1 and L1 in quadrature). $\chi^2_{\rm win}/n$ is the sum of squared whitened gated residuals over the $n$ analysed samples inside the time windows of Figs.~\ref{fig:maxl_waveforms} and~\ref{fig:maxl_waveforms_N0} ($-120$ to $+12$\,ms for the inspiral sector, $-5$ to $+15$\,ms for the ringdown sector, relative to $t_c$ of the overtone point). $\chi^2_{220}$ is the same sum over the ringdown window when only the $(2,2,0)$ mode of the template is subtracted. OT: overtone hypothesis; HM: horizon-mode hypothesis.}
\label{tab:maxl_fit}
\begin{ruledtabular}
\begin{tabular}{llcccc}
 & & \multicolumn{2}{c}{$N=+3$} & \multicolumn{2}{c}{$N=0$}\\
 & & OT & HM & OT & HM\\
\hline
Inspiral & $\rho_{\rm mf}$ & 69.43 & 69.52 & 65.43 & 64.67\\
 & $\chi^2_{\rm win}/n$, H1 & 254.1/249 & 253.7/249 & 250.6/245 & 255.5/244\\
 & $\chi^2_{\rm win}/n$, L1 & 243.8/248 & 243.1/248 & 246.0/245 & 239.5/244\\
Ringdown & $\rho_{\rm mf}$ & 35.03 & 34.96 & 37.25 & 37.59\\
 & $\chi^2_{\rm win}/n$, H1 & 33.1/28 & 33.9/28 & 37.5/31 & 35.1/32\\
 & $\chi^2_{\rm win}/n$, L1 & 25.4/29 & 22.9/29 & 26.7/31 & 26.2/32\\
 & $\chi^2_{220}$, H1 & 145.9 & 184.1 & 296.0 & 799.0\\
 & $\chi^2_{220}$, L1 & 117.3 & 253.1 & 348.7 & 794.7\\
\hline
\multicolumn{2}{l}{$\ln\mathcal{L}_{\rm max}-\ln\mathcal{L}^{\rm OT}_{\rm max}$} & 0 & 2.34 & 0 & 1.94\\
\multicolumn{2}{l}{$\ln\mathcal{B}^{\rm HM}_{\rm OT}$} & \multicolumn{2}{c}{$3.36\pm0.12$} & \multicolumn{2}{c}{$1.83\pm0.13$}\\
\end{tabular}
\end{ruledtabular}
\end{table*}

\begin{table}[t]
\caption{Difference between the maximum-likelihood points of the two hypotheses (horizon mode minus overtone) of the two terms of Eq.~\eqref{eq:lnL_split}, summed over H1 and L1, for each sector and for both sectors together. The last row is the difference of the log-likelihoods at each sector's maximum-likelihood polarisation; marginalised over polarisation, as in the sampling, the totals are $2.34$ and $1.94$.}
\label{tab:maxl_split}
\begin{ruledtabular}
\begin{tabular}{lcccccc}
 & \multicolumn{3}{c}{$N=+3$} & \multicolumn{3}{c}{$N=0$}\\
 & insp. & ringd. & both & insp. & ringd. & both\\
\hline
Data term & $-5.97$ & $+4.58$ & $-1.39$ & $+48.43$ & $-10.17$ & $+38.25$\\
Template term & $+6.02$ & $-2.52$ & $+3.50$ & $-49.29$ & $+12.56$ & $-36.73$\\
Sum & $+0.05$ & $+2.07$ & $+2.12$ & $-0.86$ & $+2.39$ & $+1.52$\\
\end{tabular}
\end{ruledtabular}
\end{table}

\section{Ringdown signal model}
\label{app:ringdown_model}

A perturbed Kerr black hole relaxes to equilibrium by radiating a
superposition of damped sinusoids, the quasinormal modes (QNMs).
Each mode is labelled by a triplet of integers: the angular
indices $\ell \ge 2$ and $-\ell \le m \le \ell$, together with an
overtone index $n = 0, 1, 2, \dots$, where $n = 0$ denotes the
fundamental mode of the multiplet and $n \ge 1$ the successive
overtones. The corresponding complex frequency can be written as
$\Omega_{\ell mn} = \omega_{\ell mn} + i/\tau_{\ell mn}$, with
$\omega_{\ell mn}$ the oscillation frequency and $\tau_{\ell mn}$
the damping time. For a Kerr black hole of mass $M_f$ and
dimensionless spin $\chi_f = J_f / M_f^2$, the values of
$\Omega_{\ell mn}$ follow from the Teukolsky equation. The no-hair
theorem then implies that the full QNM spectrum is fixed once
$(M_f, \chi_f)$ are specified, with $\chi_f \in (-1, 1)$. A mode
is called co-rotating (counter-rotating) when its pattern speed
has the same (opposite) sign as the black hole angular momentum.

\subsection{Time-domain waveform}

The \texttt{TdQNMfromFinalMassSpin} model implemented in
\texttt{PyCBC} represents the ringdown as a sum of exponentially
damped sinusoids. Using spin-weighted spherical harmonics for the
angular dependence, the two polarisations of the emitted
gravitational wave take the form
\begin{equation}
\begin{split}
    h_{+} + i h_{\times}
    = \frac{M_f}{D_L}\sum_{\ell m n}
      &{}_{-2}Y_{\ell m}(\iota, \varphi)\,A_{\ell mn}\\
      &\times e^{-t/\tau_{\ell mn}}\,
             e^{\,i(\omega_{\ell mn} t + \phi_{\ell mn})},
\end{split}
\label{eq:ringdown_sum}
\end{equation}
where $D_L$ is the luminosity distance, ${}_{-2}Y_{\ell m}$ is
the spin-weight $-2$ spherical harmonic evaluated at the
inclination $\iota$ and azimuth $\varphi$, and $A_{\ell mn}$ and
$\phi_{\ell mn}$ are the mode amplitude and phase. The latter two
are determined by the configuration of the remnant at formation
and are therefore treated as free parameters. The sum over $m$ in
Eq.~\eqref{eq:ringdown_sum} includes both the $+m$ and $-m$
contributions. Equation~\eqref{eq:ringdown_sum} uses the time dependence
$e^{i\Omega_{\ell mn}t}$, whereas the main text uses $e^{-i\omega t}$ with
$\omega=\omega_R-i\omega_I$; the two are related by
$\omega_{\ell mn}=\omega_R$ and $1/\tau_{\ell mn}=\omega_I$, together with a complex conjugation. Time is
measured from the reference time of the ringdown model, $M_f$ is the detector-frame mass, and the use of
spherical in place of spheroidal harmonics is an approximation made by the template.

Two relations connect the $\pm m$ modes of a given $(\ell, n)$
pair. The first pairs a mode with its mirror; it is not a statement that the independently labelled co- and counter-rotating branches of the Kerr spectrum have equal damping. Their complex frequencies satisfy
\begin{equation}
    \omega_{\ell\,-m\,n} = -\omega_{\ell m n}, \qquad
    \tau_{\ell\,-m\,n} = \tau_{\ell m n},
    \label{eq:pm_relation}
\end{equation}
and, whenever the initial perturbation respects reflection
symmetry about the equatorial plane of a frame aligned with the remnant spin, their amplitudes and phases
obey
\begin{equation}
    A_{\ell\,-m\,n}\, e^{i\phi_{\ell\,-m\,n}}
    = (-1)^{\ell}\,
      A_{\ell m n}\, e^{-i\phi_{\ell m n}}.
    \label{eq:reflection}
\end{equation}
Numerical-relativity waveforms show that this symmetry is
typically well preserved during the ringdown, which justifies
adopting Eq.~\eqref{eq:reflection} in the analysis. Under this
assumption the $\pm m$ pair can be merged into a single
contribution characterised by an intrinsic amplitude
$A_{\ell |m| n}$ and a phase $\phi_{\ell |m| n}$ for each
$(\ell |m| n)$ multiplet.

\subsection{Remnant-constrained frequencies and damping times}

The frequencies and damping times are evaluated at the sampled
remnant mass and spin $(M_f,\chi_f)$. The fundamental $220$ mode and,
under the overtone hypothesis, the first Kerr overtone $221$ use the Kerr
QNM spectrum. The additional component in the horizon-mode hypothesis
instead uses the horizon-frequency prescription below.
We label this component $221_{\rm H}$, with $(\ell,m,n)=(2,2,1)$.
Here the horizon-mode index $n=1$ is the ladder index denoted by $j=1$
in Eq.~\eqref{eq:matsubara}; it is distinct from the Kerr QNM overtone
index despite the shared numerical label.
In geometric units $G=c=1$, let $q_f=\sqrt{1-\chi_f^2}$. Then
\begin{equation}
 \Omega_{\rm H}=\frac{\chi_f}{2M_f(1+q_f)},\qquad
 \kappa=\frac{q_f}{2M_f(1+q_f)}.
 \label{eq:appendix_horizon_parameters}
\end{equation}
The $221_{\rm H}$ complex frequency in the main text's
$e^{-i\omega t}$ convention is
\begin{equation}
 \omega^{\rm H}_{221}=2\Omega_{\rm H}-i\kappa.
 \label{eq:appendix_horizon_frequency}
\end{equation}
For the positive-phase time-domain convention of
Eq.~\eqref{eq:ringdown_sum}, the corresponding oscillation frequency
and damping time are $\omega_{221}^{\rm HM}=2\Omega_{\rm H}$ and
$\tau_{221}^{\rm HM}=\kappa^{-1}$, so that the complex parameter in
$e^{i\Omega t}$ is $\Omega_{221}^{\rm HM}=2\Omega_{\rm H}+i\kappa$.
Thus the component has time dependence
$e^{-\kappa t}e^{i(2\Omega_{\rm H}t+\phi_{221}^{\rm HM})}$.
It is not the Kerr $222$ QNM.

For the Kerr QNM components, we obtain the frequencies and damping
times from tabulated Kerr solutions and interpolate them with a cubic
spline. The prior on the remnant parameters is the same as in
App.~\ref{app:method}: flat in $M_f\in[10,200]\,M_\odot$ and
$\chi_f\in[-0.99,0.99]$. The amplitudes and phases of the retained
components are free parameters, with the priors of App.~\ref{app:method}.

\subsection{Competing mode content}
\label{app:ringdown_model_modes}

Both models contain the fundamental Kerr QNM
$(\ell,m,n)=(2,2,0)$, with free amplitude $A_{220}$ and phase
$\phi_{220}$. The two hypotheses differ in their second component:
\begin{itemize}
 \item In the \textbf{horizon-mode hypothesis}, the second component
 is the $(\ell,m,n)=(2,2,1)$ horizon mode, denoted $221_{\rm H}$,
 with free amplitude $A^{\rm HM}_{221}$ and phase $\phi^{\rm HM}_{221}$.
 Its frequency and damping time are fixed by
 Eqs.~\eqref{eq:appendix_horizon_parameters}--\eqref{eq:appendix_horizon_frequency}
 at the sampled remnant parameters. The Kerr $221$ overtone amplitude
 is fixed to zero.
 \item In the \textbf{overtone hypothesis}, the second component is
 the first Kerr QNM overtone $(\ell,m,n)=(2,2,1)$, with free amplitude
 $A_{221}$ and phase $\phi_{221}$. The horizon-mode amplitude
 $A^{\rm HM}_{221}$ is fixed to zero.
\end{itemize}
We parameterize the second-component phases relative to the fundamental:
\begin{equation}
\begin{aligned}
 \phi_{221}&=\phi_{220}+\Delta\phi_{221},\\
 \phi^{\rm HM}_{221}&=\phi_{220}+\Delta\phi^{\rm HM}_{221}.
\end{aligned}
\label{eq:appendix_relative_phases}
\end{equation}
This specifies a common reference phase and a relative phase in each
hypothesis. It is a change of parameters and does not by itself remove any degeneracy. The Bayes factor compares these two frequency prescriptions
with their adopted amplitude and phase priors; it does not by itself
identify the physical origin of the second component.

\subsection{Relation to the numerical-relativity literature}

The Kerr-QNM part of Eq.~\eqref{eq:ringdown_sum} is the
standard remnant-parameterized ringdown model used in spectroscopy.
The horizon-mode hypothesis supplements its $220$ component with a
damped sinusoid at the horizon frequency
$\omega^{\rm H}_{221}=2\Omega_{\rm H}-i\kappa$.
Evidence for Kerr overtone excitation in numerical-relativity waveforms
does not by itself establish a separate radiative component at this
horizon frequency. The interpretation of the horizon-mode template,
including the distinction between a useful finite-time template and a
surviving asymptotic horizon-mode contribution, is discussed in
Sec.~\ref{sec:forced}.
The mode content of numerical waveforms and its angular decomposition
are discussed in Refs.~\citep{Giesler:2019uxc,Borhanian:2019kxt}, while the
reflection-symmetry assumption is discussed in
Refs.~\citep{Berti2007,Capano2021}.

\bibliography{main}

\end{document}